\documentclass[numsec,webpdf,modern,large]{oup-authoring-template}
\usepackage{ifthen}
\newboolean{arxiv_version}
\setboolean{arxiv_version}{true}
\usepackage{dsfont}
\usepackage{physics}
\usepackage{soul}
\usepackage{algorithm}
\usepackage{algpseudocode}
\usepackage{algorithm,algcompatible,amsmath}

\algnewcommand\INPUT{\item[\textbf{Input:}]}
\algnewcommand\OUTPUT{\item[\textbf{Output:}]}
\usepackage{rotating}
\usepackage{graphicx}
\usepackage{hyperref}
\hypersetup{colorlinks,
citecolor=blue,
linkcolor=blue
}
\usepackage{bbm}
\definecolor{mygreen}{HTML}{008000}
\definecolor{mypink}{HTML}{BF00BF}
\definecolor{myblue}{HTML}{1010FF}
\definecolor{mygrey}{HTML}{BBBBBC}

\begin{document}
\ifthenelse{\boolean{arxiv_version}}{\journaltitle{Pre-print}}{\journaltitle{Submission to Bioinformatics}}
\copyrightyear{2025}
\pubyear{2025}
\firstpage{1}

\title[Topological model selection]{Topological model selection: a case-study in tumour-induced angiogenesis}
\subtitle{}
\author[1]{Robert A McDonald}
\author[1,2]{Helen M Byrne}
\author[1,3,4,5]{Heather A Harrington} 
\author[6,$\ast$]{Thomas Thorne}
\author[7,8,$\ast$]{Bernadette J Stolz}

\authormark{McDonald et al.}
\address[1]
{Mathematical Institute, 
University of Oxford, 
Radcliffe Observatory Quarter, Woodstock Road,
Oxford, OX2 6GG, 
United Kingdom}
\address[2]
{Ludwig Institute for Cancer Research,
Nuffield Department of Medicine,
Old Road Campus Research Building, 
Oxford OX3 7DQ,
United Kingdom}
\address[3]
{Faculty of Mathematics,
Technische Universitat Dresden,
01062 Dresden,
Germany}
\address[4]
{Centre for Systems Biology Dresden (CSBD),
Pfotenhauerstrasse 108,
01062 Dresden,
Germany}
\address[5]
{{Max Planck Institute of Molecular Cell Biology and Genetics (MPI-CBG),
01307 Dresden},
Germany}
\address[6]
{Computer Science Research Centre,
University of Surrey,
GU2 7XH,
United Kingdom}
\address[7]
{Department of Machine Learning and Systems Biology,
Max Planck Institute of Biochemistry, Am Klopferspitz 18,
82152
Martinsried,
Germany}
\address[8]{Munich Center for Machine Learning,
Oettingenstraße 67,
80538 Munich, Germany
}
\corresp[$\ast$]{Corresponding authors. 
\href{mailto:tom.thorne@surrey.ac.uk}{tom.thorne@surrey.ac.uk},
\href{mailto:stolz@biochem.mpg.de}{stolz@biochem.mpg.de}}

\abstract{
\textbf{Motivation:}
Comparing mathematical models offers a means to evaluate competing scientific theories.
However, exact methods of model calibration are not applicable to many probabilistic models which simulate high-dimensional spatio-temporal data.
Approximate Bayesian Computation is a widely-used method for  parameter inference and model selection in such scenarios, and
it may be combined with Topological Data Analysis to study models which simulate data with fine spatial structure.\\
\textbf{Results:}
We develop a flexible pipeline for parameter inference and model selection in spatio-temporal models.
Our pipeline identifies topological summary statistics which quantify spatio-temporal data and uses them to approximate parameter and model posterior distributions.
We validate our pipeline on models of tumour-induced angiogenesis, inferring four parameters in three established models and identifying the correct model in synthetic test-cases.
\\
\textbf{Availability and implementation:} Simulation code for all models, data analyses, parameter inference and model selection is available online at \url{https://github.com/rmcdomaths/tms/} and archived at \url{https://doi.org/10.5281/zenodo.17392787}. \\
\textbf{Supplementary information:} \textit{Supplementary Information} will be available online.}

\maketitle

\section{Introduction}
Given multiple mathematical models which aim to reproduce the same biological data, determining which model and parameters give the best fit presents a theoretical and computational challenge. 
For example, spatio-temporal models often simulate complex high-dimensional data which is difficult to quantify and compare to observed data.
Such models do not in general yield tractable likelihood functions, which significantly hinders the use of exact methods for parameter inference and model selection \citep{modelSelectionReview}. 

Many mathematical models have been developed to study the mechanisms underlying tumour-induced angiogenesis \citep{preziosireview, vilanovareview}, a hallmark of cancer \citep{hallmarks}.
Tumour cells use chemical signals to stimulate the growth of new blood vessels from existing vasculature \citep{VEGFforAngio}, which provide a tumour mass with oxygen and nutrients that it requires to grow.
However, instead of concise equations determining the growth of angiogenic networks, such models often comprise multiple agents and heterogeneous environments whose interactions depend non-deterministically on their spatial organisation.
Discrete models of tumour-induced angiogenesis, for example, use multiple model rules and parameters to reproduce the branches, loops, and multiple components that characterise real vascular networks.

We use Topological Data Analysis (TDA), Approximate Bayesian Computation (ABC), and Random Forests (RFs) to develop a pipeline for parameter inference and model selection applicable to spatio-temporal models.
TDA offers a toolkit of methods for quantifying spatial data \citep{ghristBarcodes, carlssonTDA, edelsbrunnerHarer}.
TDA has previously been used to study multi-agent temporal systems \citep{tdaBiologicalAggregation, TDAcollectivemotion, relationalPH} and was used in related work to compare models of insect locomotion \citep{TDAbiomodels} and pattern formation in zebrafish \citep{sandstedeModelTDA}.
ABC provides a statistical framework for using model simulations to approximate posterior distributions when likelihood functions are not available \citep{ABCreview}.
RFs are an ensemble estimation method from machine learning which have previously been combined with ABC to estimate parameter values \citep{ABCRF_paraminf} and rank candidate models \citep{ABCRF_modelchoice}.

We begin by outlining three existing models of tumour-induced angiogenesis in which exact methods of parameter inference and model selection are not applicable.
We show how TDA can be used to characterise spatial data simulated by the models and briefly describe methods from ABC and RF.
We then present a three-step pipeline for parameter inference and model selection which we apply to the angiogenesis models.
Commenting on the applicability of our pipeline to experimental data, we discuss how topological summaries may be used to evaluate a variety of modelling approaches in biology.

\section{Model data and analysis}
\subsection{Angiogenesis models}\label{sec:angio}
Discrete models of tumour-induced angiogenesis simulate the movement of individual Endothelial Cells (ECs).
Vascular Endothelial Growth Factors (VEGF) produced by tumour cells initiate a cascade of chemical reactions which drive ECs towards the tumour.
Early models employed the \textit{snail-trail model} \citep{originalSnailTrail} in which tip ECs migrate up spatial gradients of VEGF and fibronectin, while stalk ECs proliferate in their path to produce a contiguous line of cells.
When two separate trails of ECs meet they fuse together to form a loop, in a process known as anastomosis.
A tip EC may also split into two tip ECs, which thereafter move independently.
Migrating, branching and looping tip ECs eventually reach the tumour mass and the connecting trails of stalk ECs form a blood vessel network.
Recent models reflect modern discoveries of cell mixing and phenotype switching \citep{angioCellMixing}, where ECs change type and overtake each other before forming a stable blood vessel network.
Other models view ECs as a continuous population density rather than individual cells \citep{continuumangiomodels} and account for blood flow and nutrient delivery when simulating vascular networks \citep{flowangiomodels}.
We develop our pipeline of parameter inference and model selection on discrete angiogenesis models due to their simulation of finely resolved spatial data.

The Anderson-Chaplain (AC) \citep{andersonChaplain}, Stokes-Lauffenberger (SL) \citep{stokesLauffenberger}, and Plank-Sleeman (PS) \citep{plank_sleeman} models employ the snail-trail model to simulate movement of individual ECs in a two-dimensional, square domain.
We assume that VEGF levels increase from the bottom of the domain to a tumour at the top, guiding ECs to move upwards.
Each model initalises multiple distinct tip ECs
along the bottom of the domain, simulating their trajectories according to model-specific movement rules.
In each model, we choose four parameters that are likely to lead to measurable changes in simulated data and we attempt to infer their values. 
Figure \ref{fig:importance} illustrates the movement rules and model parameters in each model.
See \textit{Supplementary Information} section \ifthenelse{\boolean{arxiv_version}}{\ref{SI:sec:models}}{1}
for full statements of each model and its parameters.

\begin{figure}[t]
\includegraphics[width=.45\textwidth]{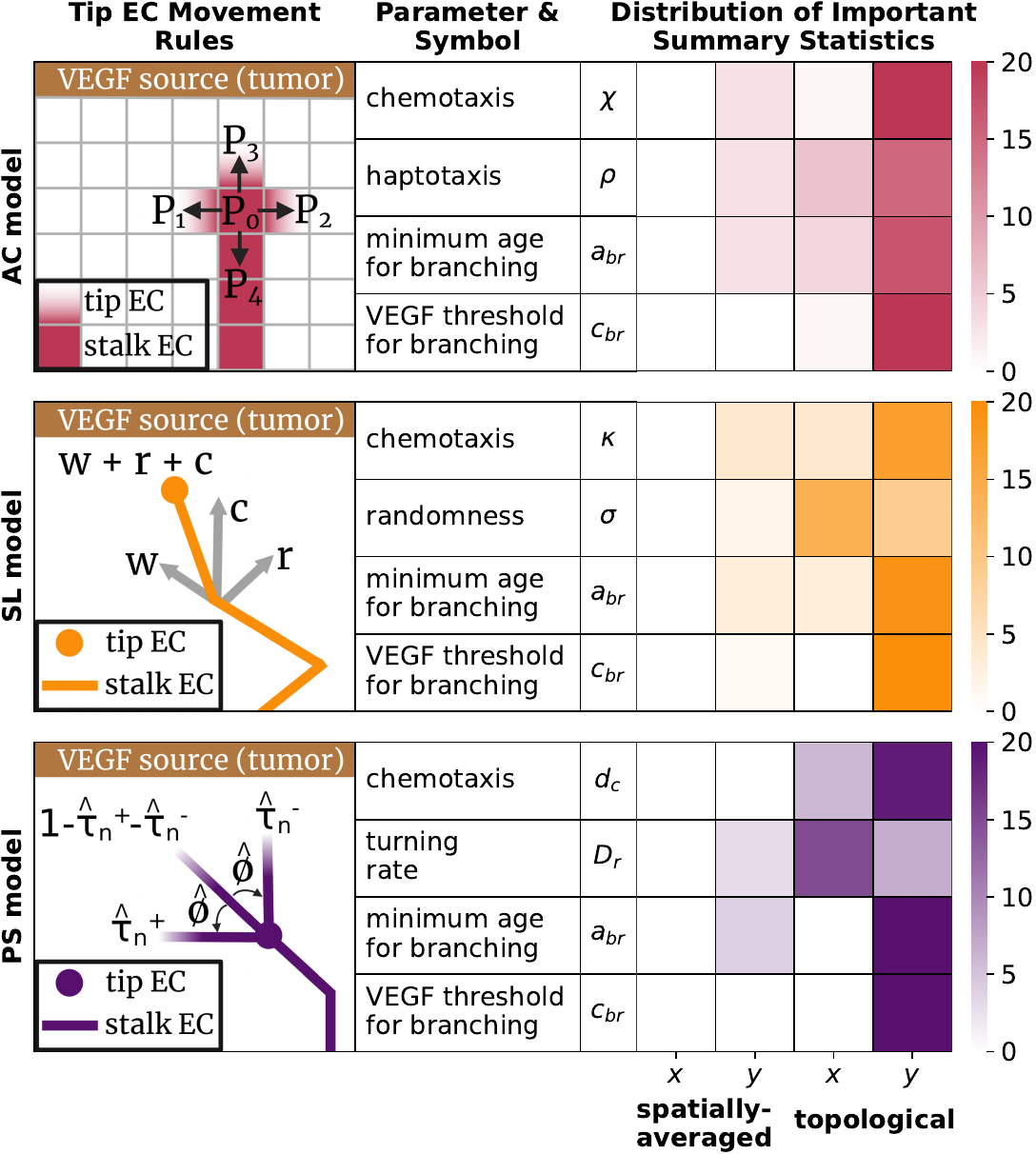}
\caption{
In the AC model \citep{andersonChaplain}, a tip endothelial cell (EC) makes one of five possible moves on a square lattice in each time-step according to probabilities $P_0, P_1, P_2, P_3, P_4$.
A chemotaxis parameter $\chi$ biases movement probabilities in the direction of increasing VEGF concentration, and a haptotaxis parameter $\rho$ biases moves in the direction of increasing fibronectin.
In the SL model \citep{stokesLauffenberger} tip ECs move in any direction (off-lattice) with velocities modelled by a two-dimensional stochastic differential equation.
Parameters $\kappa$ and $\sigma$ determine how strongly an EC's current velocity $w$ is affected by the VEGF gradient $c$, and random variation $r$ respectively. 
The PS model \citep{plank_sleeman} assigns a constant speed to each tip EC and, at each time-step, rotates the angle that the velocity vector makes with the vertical.
The probability $\hat{\tau}_n^+ + \hat{\tau}_n^-$ that a tip EC turns by $\hat{\phi}$ is determined by a turning rate parameter $D_r$. 
A chemotaxis parameter $d_c$ biases turns that re-orient the EC's direction  
towards the tumour.
In all models, a tip EC may bifurcate into two ECs which thereafter move independently 
if its age exceeds
the minimum age for branching parameter $a_\text{br}$ and the VEGF concentration at its location exceeds
the the VEGF threshold for branching parameter $c_\text{br}$.
We show how many spatially-averaged and topological summary statistics, computed in either the $x$ or $y$ co-ordinate direction, appear among the $100$ most important summary statistics to the inference of each parameter.
} \label{fig:importance}
\end{figure}

\subsection{Data Generation and Analysis}\label{sec:dataAnalysis}
Each angiogenesis model outlined in section \ref{sec:angio} (and described fully in \textit{Supplementary Information} section \ifthenelse{\boolean{arxiv_version}}{\ref{SI:sec:models}}{1}) simulates EC movement in a square domain. 
To summarise the spatial properties of each simulation, we overlay a regular grid onto the domain at the final timestep and compute a collection of \textit{spatially-averaged} and \textit{topological} summary statistics.
For the \textit{spatially-averaged} summary statistics, we compute the mean, standard deviation, minimum, maximum, range, and the 10th, 25th, 75th and 90th percentiles of the $x$ and $y$ co-ordinates of EC locations in the grid.
These were used in \cite{nardiniclustering} to distinguish the AC model's behaviour in different parameter regimes.

 Persistent homology (PH) is a prominent method within Topological Data Analysis (TDA) \citep{ghristBarcodes, carlssonTDA, edelsbrunnerHarer} to quantify loops, branches, and connected components. 
 Here, we require finer information than is provided by standard persistence; therefore, we use extended persistent homology (EPH) \citep{extendedpersistence}.
 We give an overview of PH and EPH, briefly indicating how EPH arises from PH and captures a greater range of spatial information within the angiogenesis data we simulate. 
 See \textit{Supplementary Information} section \ifthenelse{\boolean{arxiv_version}}{\ref{SI:sec:ss}}{2} for a full definition of EPH and a worked example.
\begin{figure}[b]
    \centering
    \includegraphics[width=.45\textwidth]{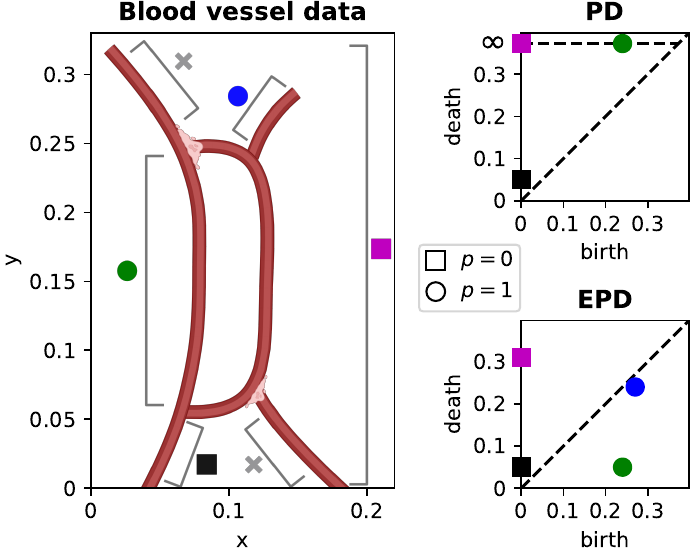}
    \caption{A persistence diagram (PD) and extended persistence diagram (EPD) for a simple blood vessel computed using the vertical sweeping-plane filtration. 
    The PD points quantify the size and location of the small lower branch (\scalebox{1}{\color{black}$\blacksquare$}), and the locations of the component (\scalebox{1}{\color{mypink}$\blacksquare$}) and loop (\scalebox{1.75}{\color{mygreen}$\bullet$}).
    The EPD points quantify the location \textit{and size} of all topological features quantified by the PD, in addition to the small upper branch (\scalebox{1.75}{\color{myblue}$\bullet$}).
    The branches ({\protect\scalebox{.3}{\protect\textcolor{mygrey}{\protect\tikz{\protect\draw[line width=5pt] (0,0) -- (0.5,0.5) (0,0.5) -- (0.5,0);}}}}) are not detected by PH or EPH with this sweeping-plane filtration.
}
    \label{fig:ph_eph}
\end{figure}

To compute PH, a nested sequence of simplicial complexes, known as a filtration, is built on the data. 
Intuitively, a simplicial complex $\Sigma_k$ is a graph that includes nodes and edges as well as higher-order connections such as triangles or tetrahedra.
A filtration must be carefully constructed such that each $\Sigma_k$ encodes the spatial properties of the underlying data at some spatial threshold defined by $k$. 
For example, \cite{nardiniclustering} used a \textit{sweeping-plane} filtration to analyse the AC model, where $\Sigma_k$ is constructed from those ECs which are a distance of $k$ or more away from the tumour.
Once a filtration has been chosen, one computes a sequence of $\mathbb{F}$-vector spaces $H_p(\Sigma_k)$ known as homology groups.
We use the field $\mathbb{F} = \mathbb{Z} / 2\mathbb{Z}$, which is widely adopted in
applications for its simplicity and interpretability. 
Homology groups quantify $p$-dimensional topological features in each $\Sigma_k$ (see, for example, \cite{roadmap}).
$H_0$ detects connected components, $H_1$ detects loops and, in general, $H_p$ detects $p$-dimensional voids. 
We consider dimensions $0$ and $1$ only, since voids of dimension $2$ or higher do not appear in the angiogenesis data we simulate.
Persistence pairs $(b, d)$ are computed from the sequence of homology groups to quantify topological features in the filtration \citep{structurePH}.
A birth $b$ corresponds to the index $k$ in the filtration at which a topological feature first appears.
A death $d$ is either the index $k$ at which the $p$-dimensional void it represents is filled in, or $\infty$ if the topological feature persists through the filtration.
The difference $d-b$ is known as the persistence of a topological feature.
The interpretation of $(b, d)$ in terms of the underlying data depends on the choice of filtration.
In the sweeping-plane filtration used in \cite{nardiniclustering}, persistence pairs quantify connected components and loops in simulated vascular networks in terms of their distance from the tumour.
 
In PH, some topological features typically persist throughout the entire filtration.
In Figure \ref{fig:ph_eph}, for example, a connected component appears at $y=0$ and a loop at $y=0.25$, and both persist for all values of $y$ in the sweeping-plane filtration.
The corresponding persistence pairs are therefore $(0, \infty)$ and $(0.25, \infty)$, which quantify limited location information and no size information about the topological features they represent.
Furthermore, PH computes one PD in each dimension ($p=0$ and $p=1$ in this work), but we want to distinguish several different spatial structures (see Figure \ref{fig:ph_eph} for examples).

We would like to quantify the size and location of different spatial features. EPH provides this information by appending \textit{relative} homology groups \citep{edelsbrunnerHarer} to the sequence of ordinary homology groups in PH.
Topological features which persist through all ordinary homology groups will die in the relative homology groups, so persistence pairs computed from EPH are guaranteed to have finite persistence.
Each EPH persistence pair then be classified as one of four types depending on where the birth and death appear in the sequence of ordinary and relative homology groups, which provides additional information about the corresponding topological features.
See \textit{Supplementary Information} section \ifthenelse{\boolean{arxiv_version}}{\ref{SI:sec:ss}}{2} for a formal definition of topological feature types in EPH and their interpretation in simulated angiogenesis data.
Figure \ref{fig:ph_eph} compares persistence pairs computed from a simple blood vessel network using PH and EPH, illustrating the extra information provided by EPH.
We compute two extended persistence diagrams (EPDs) for each angiogenesis dataset--using a vertical $(y)$ and a horizontal $(x)$ sweeping-plane filtration. 
We vectorise each EPD using Persistence Images \citep{persistenceimage} and 
persistence statistics \citep{vectorisationSurvey}, and our \textit{topological} summary statistics are the concatenation of these vectors.
% --see \textit{Supplementary Information} section \ifthenelse{\boolean{arxiv_version}}{\ref{SI:sec:ss}}{2} for full definitions.

\section{Methods}
\subsection{Approximate Bayesian Computation}\label{subsec:ABC}
ABC provides a statistical framework for using data to infer model parameters.
Suppose a model uses parameters $\Theta$ to simulate data $\mathcal{D}$ according to some probability distribution $p(\mathcal{D} | \Theta)$, called the \textit{likelihood}.
Parameter inference aims to determine the \textit{posterior} distribution $p(\Theta | \mathcal{D})$, which is the probability that parameters $\Theta$ generated observed data $\mathcal{D}$.
Using previous experiments or assumptions about feasible parameter values, one may define a \textit{prior} distribution $p(\Theta)$ representing knowledge of the parameter values before data has been observed.
The likelihood, prior and posterior are related by Bayes' rule,
\begin{equation}
   p(\Theta | \mathcal{D}) = \frac{p(\mathcal{D} | \Theta) p(\Theta)}{p(\mathcal{D})},\nonumber
\end{equation}
where the evidence $p(\mathcal{D})$ is the integral $\int_\Theta p(\mathcal{D} | \Theta) p(\Theta) \textnormal{d} \Theta$ over parameters $\Theta$ in the support of the prior.
Although Bayes' rule gives a closed formula for the posterior distribution, it is often impractical to use directly.
The likelihood function $p(\mathcal{D} | \Theta)$ may be too complicated to derive for probabilistic spatial models in which many datasets $\mathcal{D}$ may be simulated from the same parameters $\Theta$.
Instead, Bayes' rule is used to derive Approximate Bayesian Computation (ABC) algorithms \citep{ABCreview} which allow sampling from the posterior when the likelihood and evidence are not known.
ABC algorithms sample candidate parameters $\theta_i$ from the prior $p(\theta)$ and accept them if the distance $\nu(\mathcal{D}_i, \mathcal{D}^*)$ between simulated and observed data is less than some tolerance $\epsilon > 0$ for some distance function $\nu$.
If the tolerance $\epsilon$ is set to zero, then the distribution of accepted parameters is the posterior $p(\Theta | \mathcal{D}^*)$ \citep{ABCproof}.
However, it is often not appropriate or possible to seek an exact posterior distribution from observed data, since a model may rarely reproduce observed data $\mathcal{D^*}$ exactly, and the observed data may be noisy.
It is therefore advisable to choose $\nu$ and $\epsilon$ such that parameter values are accepted if they simulate data that is similar to observed data.
The general form of such a distance function is $\nu (\mathcal{D}^*, \mathcal{D}_i) = \norm{X^* - X_i}_2$ where $X_i$ and $X^*$ are vectors of \textit{summary statistics} computed from model data $\mathcal{D}_i$ and observed data $\mathcal{D}^*$ respectively.
Summary statistics aim to capture relevant properties of data as a low dimensional vector.
As $\epsilon$ approaches $0$, the distribution of parameters accepted by ABC algorithms approaches $p(\Theta | X^*)$, which equals $p(\Theta | \mathcal{D}^*)$ if the summary statistics are sufficient for the model in question, or is a close approximation if the summary statistics are insufficient but informative \citep{sufficientStats}.

\subsection{Random Forests}\label{sec:RF}
Random Forests (RFs) \citep{orginalRF} learn relationships between feature vectors and response variables.
Training data comprising a collection of feature vectors $X_i \in \mathcal{X}$ and corresponding response variables $y_i \in \mathcal{Y}$ are used to train a RF, enabling it to predict
the true response variable $y^*$ of an unseen feature vector $X^*$.
Regression RFs are used when $y_i$ are continuous values and classification RFs are used when $y_i$ are discrete labels.
\cite{ABCRF_paraminf} used a regression RF for parameter inference by using simulated data $\mathcal{D}_i$ to learn the relationship between summary statistics $X_i$ and parameter values $y_i = \theta_i$.
Given summary statistics $X^*$ of unseen data $\mathcal{D}^*$, the prediction $RF(X^*)$ predicts the true parameter value $\theta^*$.
In addition to predicting unseen feature vectors, a trained RF provides useful information about the training data.
The out-of-bag prediction $RF_\text{oob}(X_i)$ estimates the (known) response variable $y_i$ using pairs from the training data other than $(X_i, y_i)$.
The out-of-bag error rate $p(RF(X_i) \neq y_i)$ then gives a (unbiased) measure of how well the relationship between $X_i$ and $y_i$ is captured by the rest of the training data. 
A trained RF also gives a measure of the \textit{importance} of each co-variate $j$ within feature vectors $X_i = (X_i^0, \dots, X_i^j, \dots, X_i^{n_\text{f}})$ to the problem of predicting response variable $y_i$.
Intuitively, important features are those whose values within $X_i$ and $X_i'$ differ when $y_i$ and $y_i'$ do, and which are hence useful in learning the relationship between training data $\mathcal{X}$ and $\mathcal{Y}$.
\subsection{Model Selection}
Given observed data $\mathcal{D}^*$, the model posterior $p(m | \mathcal{D}^*)$ gives the probability that models $m=m_i$ generated $\mathcal{D}^*$.
ABC algorithms rely on the approximation $p(\Theta|X^*) \approx p(\Theta|\mathcal{D}^*)$, which holds as long as the vector $X_i$ carries a similar amount of information about the parameter value $\theta_i$ as the simulated data $\mathcal{D}_i$ itself.
However, the information loss suffered by a collection of summary statistics may vary between models \citep{noConfidenceABCModelChoice}, so it is inadvisable simply to infer $m_i$ as a (discrete) parameter using an ABC algorithm.
\cite{ABCRF_modelchoice} instead used two RFs to approximate $p(m|\mathcal{D}^*)$.
A classification RF learns the relationship between simulated data $X_i$ and model label $y_i = m_i$ and gives a prediction $RF(X^*)$ of the model $m^*$ which generated unseen data $\mathcal{D}^*$.
A regression RF is then trained to learn the relationship between $X_i$ and $p(RF_\text{oob}(X_i) \neq m_i)$--the out-of-bag error rate of the classification RF. 
The regression RF is then used to estimate posterior probability $p(m = m^* | \mathcal{D}^*)$ as $1 - p(RF(X^*) \neq m^*)$.

\begin{figure*}[!ht]
\includegraphics[width=.315\textwidth]{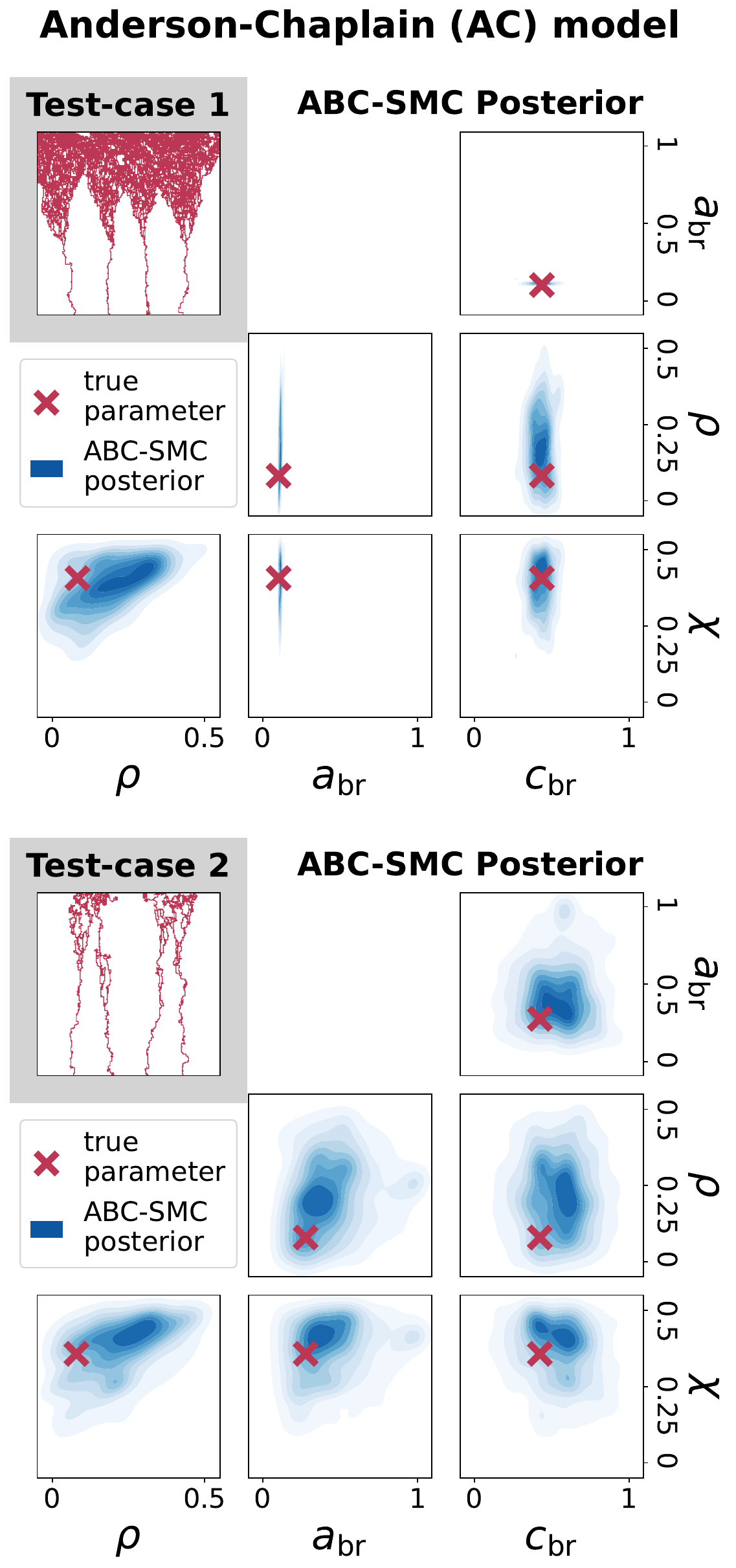}\hfill
\includegraphics[width=.315\textwidth]{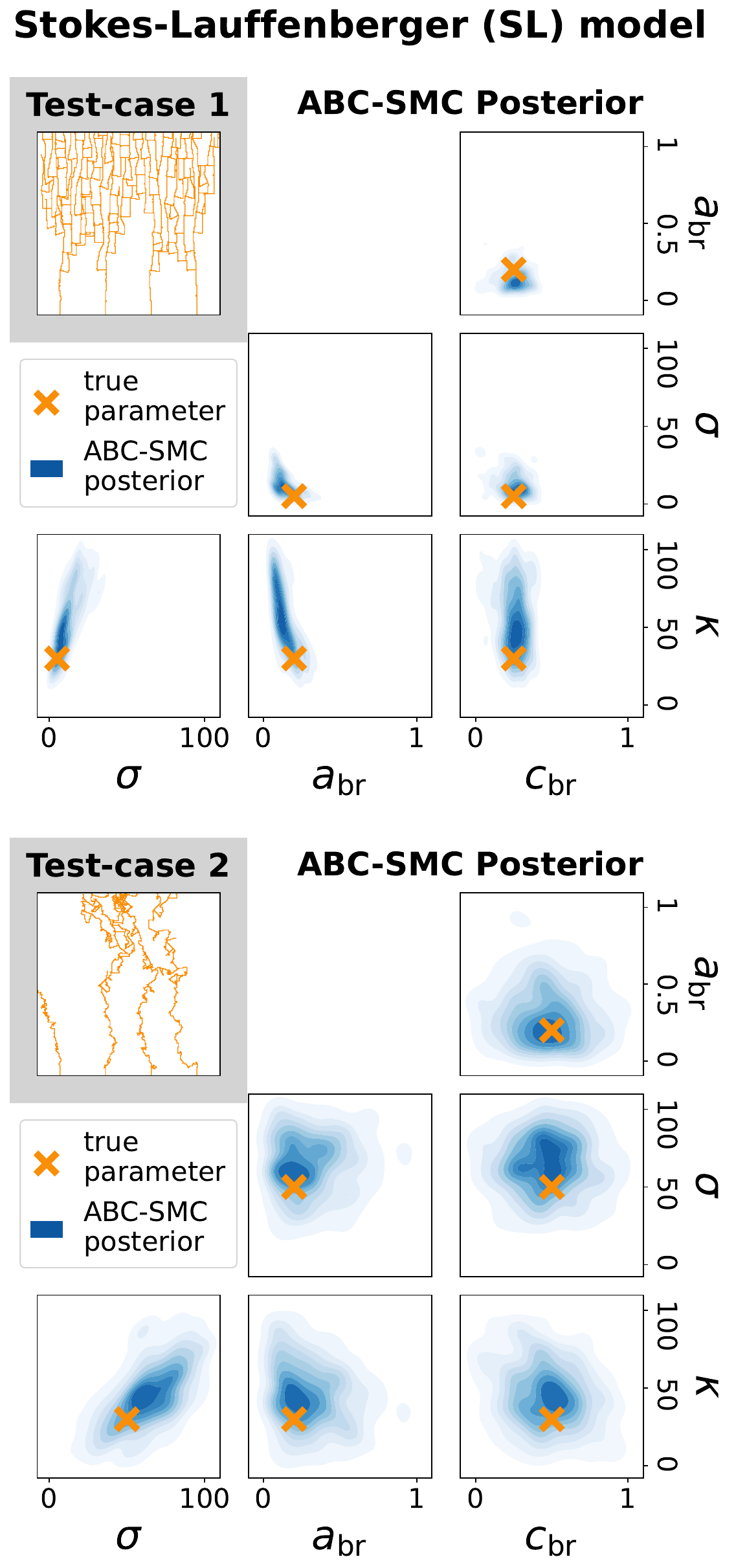}\hfill
\includegraphics[width=.315\textwidth]{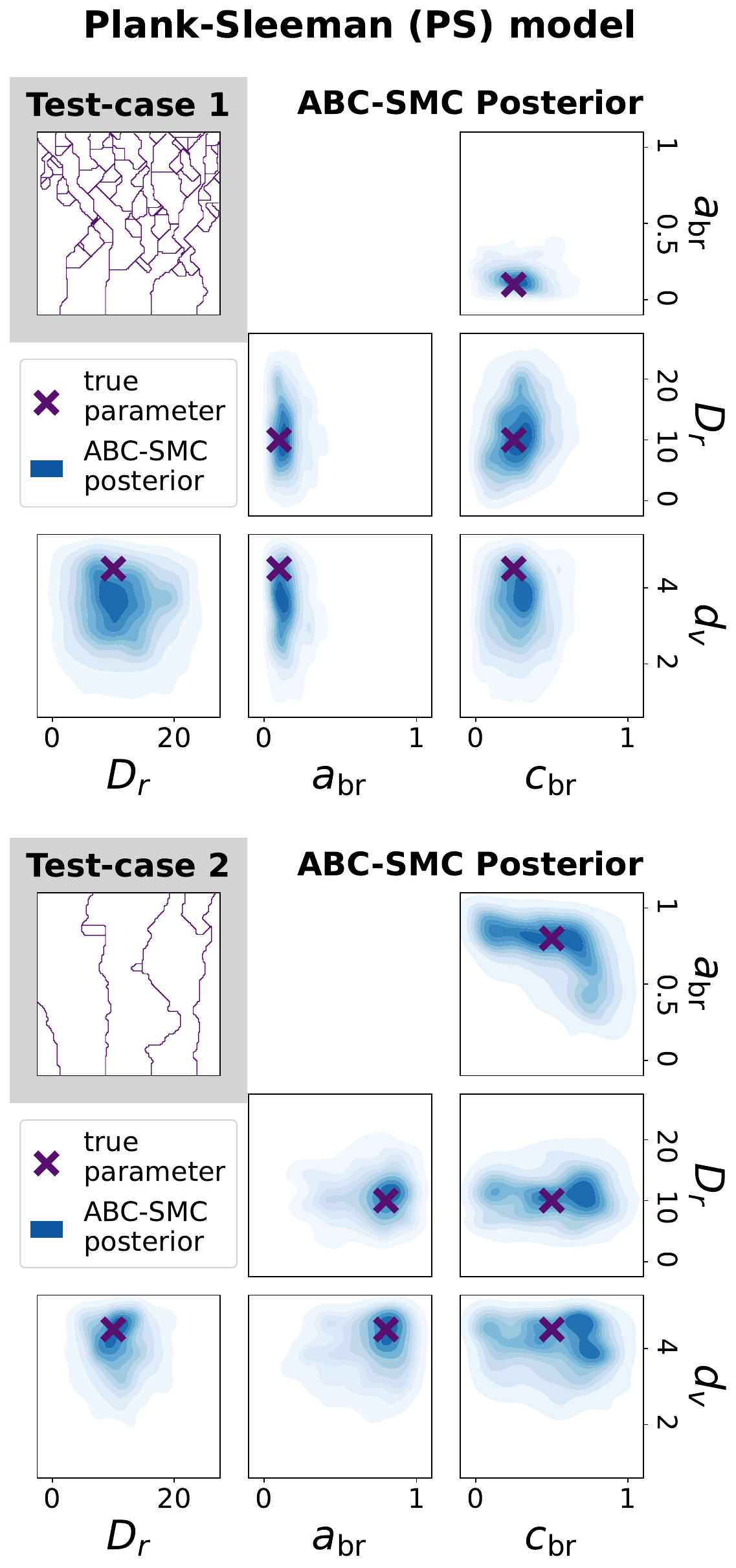}
\caption{
We infer the minimum age for branching ($a_\text{br}$) and VEGF threshold for branching ($c_\text{br}$) in each model, as well as chemotaxis and haptotaxis parameters ($\chi$ and $\rho$) in the AC model, chemotaxis and randomness parameters ($\kappa$ and $\sigma$) in the SL model, and chemotaxis and turning rate parameters ($d_c$ and $D_r$) in the PS model.
We simulate each model $10$ times at known parameter values to generate two synthetic test-cases for each model, and show the final time-step of one such simulation.
We then use steps 1-2 of section 3 to approximate the parameter posterior $p(\Theta | \mathcal{D}^*)$ in each test-case. 
We project the approximate ABC-SMC posterior to each parameter pair and plot the resulting distributions (fitting a Gaussian kernel to the parameter values accepted in the final population of the ABC-SMC algorithm), along with the true parameter which generated the test-case.
}
\label{fig:param_inference}
\end{figure*}

\section{Spatial Parameter Inference and Model Selection}\label{sec:ourmethod}
Given observed data $\mathcal{D}^*$, we wish to approximate the parameter posterior $p(\Theta|\mathcal{D}^*)$ for candidate models $m=m_1, m_2, \dots$ and the model posterior $p(m|\mathcal{D^*})$.
Informative summary statistics may be used to infer parameter values using ABC, but uninformative or poorly scaled summary statistics may misrepresent the difference between datasets generated by similar parameters \citep{summaryStatisticChoice}.
We therefore seek a collection of summary statistics that quantify simulated data and, in particular, quantify how simulated data changes when different model rules and parameters are used to generate it.
We use informative summary statistics to approximate parameter and model posteriors in a three-step pipeline.
We test this pipeline on toy models in \textit{Supplementary Information} section \ifthenelse{\boolean{arxiv_version}}{\ref{SI:sec:toymodelexample}}{4}
and apply it to the three angiogenesis models in section \ref{sec:results}.
\\

\noindent\textbf{Step 1: Identify informative summary statistics}\\
\noindent
We use RFs to find a small subset of summary statistics to be used in ABC. 
To generate training data, we draw parameter values $\theta_i$ from the prior distribution $p(\Theta)$ for each parameter in each model, simulate model data $\mathcal{D}_i$, and compute spatially-averaged and topological summary statistics $X_i$ from the final simulated time-step.
We train regression RFs to learn the relationship between summary statistics $X_i$ and parameter values $\theta_i$--one RF for each parameter in each model.
We then rank the spatially-averaged and topological summary statistics by their importance according to the RF (see section \ref{sec:RF}).
In each RF, feature importance decreases exponentially (as in \cite{ABCRF_paraminf}) and a small subset of summary statistics provides most of the predictive power of each RF.
We select an equal number of informative summary statistics from each RF, collecting a total of $n_\text{s}=100$ for each model.
See \textit{Supplementary Information} section \ifthenelse{\boolean{arxiv_version}}{\ref{SI:sec:RF}}{3}
for a full definition of RF feature importance and a discussion of how we choose 
$n_\text{s}$.
RFs identify which summary statistics quantify the effect of each parameter on simulated data and allow us to omit those summary statistics which do not. 
\\ \newline
\noindent \textbf{Step 2: Fit each model to the observed data}\\
We use the summary statistics identified by step 1 to define a distance function for use in ABC.
We use $\nu (\mathcal{D}^*, \mathcal{D}_i) = \norm{x^* - x_i}_2$, where $x_i$ is the vector $X_i$ restricted to the top $n_\text{s}$ summary statistics identified in step 1, $x^*$ is computed from observed data, and the distance is averaged over multiple instances of observed data.
We scale each summary statistic by the largest absolute value of that summary statistic in the training data.
We then use the ABC-SMC algorithm of \cite{smcDelMoral} to approximate $p(\Theta|\mathcal{D}^*)$ for each model.
By using only those summary statistics which quantify the effect of parameter values on simulated data, we ensure $\nu$ is informative about the value of $\theta$ used to generate the observed data.
Scaling ensures that each summary statistic contributes approximately equally to the distance function $\nu$ and limits the influence of poorly scaled summary statistics.\\
\\
\noindent\textbf{Step 3: Approximate the model posterior}

\noindent
Using summary statistics which are informative for all models, we use two more RFs to estimate the model posterior.
Following \cite{ABCRF_modelchoice}, we train a classification RF to learn the relationship between (unscaled) summary statistics $X_i$ and model indices $m_i$ in the training data.
We modify $X_i$ to contain only those summary statistics which appear among the $n_\text{s}$ most important summary statistics for all models.
We then train a regression RF to learn the relationship between $X_i$ and $p(RF_\text{oob}(X_i) \neq m_i)$--the probability that the predicted model index is incorrect.
The classification RF gives an estimate $RF(X^*)$ of the model $m^*$ that generated the observed data $\mathcal{D}^*$, and the regression RF is used to estimate $p(m = m^* | \mathcal{D}^*)$ as $1 - p(RF_\text{oob}(X^*) \neq m^*)$.
We choose the value of $n_\text{s}$ in step 2 large enough to ensure that some informative summary statistics are selected for all three models under consideration--we use these to approximate the model posterior.

\section{Results}\label{sec:results}

\subsection{RFs find small subsets of informative summary statistics}

We sample $n=10,000$ model parameters from uniform priors with ranges taken from existing literature, or by analysing each model's data generation rules (See \textit{Supplementary Information} section \ifthenelse{\boolean{arxiv_version}}{\ref{SI:sec:models}}{1}
for details).  
We select $n_\text{s}=100$ summary statistics for each model and report the type of summary statistics selected--spatially-averaged or topological, computed in the horizontal or vertical direction--in Figure \ref{fig:importance}.
A mixture of spatially-averaged and topological summary statistics are selected for each parameter, however there is a clear preference for topological summary statistics for each parameter.
In general, summary statistics computed in the vertical direction are selected more often than those computed in the horizontal direction, which is unsurprising, since most parameters regulate the movement of ECs upwards.
However, the randomness parameter $\sigma$ of the SL model and the turning coefficient $D_r$ of the PS model are two exceptions.
These parameters cause ECs to deviate from their upward trajectory, and their inference therefore relies on topological summary statistics computed in the horizontal direction.
No spatially-averaged summary statistics computed in the horizontal ($x$) direction are chosen, indicating that these measures are too coarse to distinguish different model simulations.
\begin{figure}[t]
\includegraphics[width=.48\textwidth]{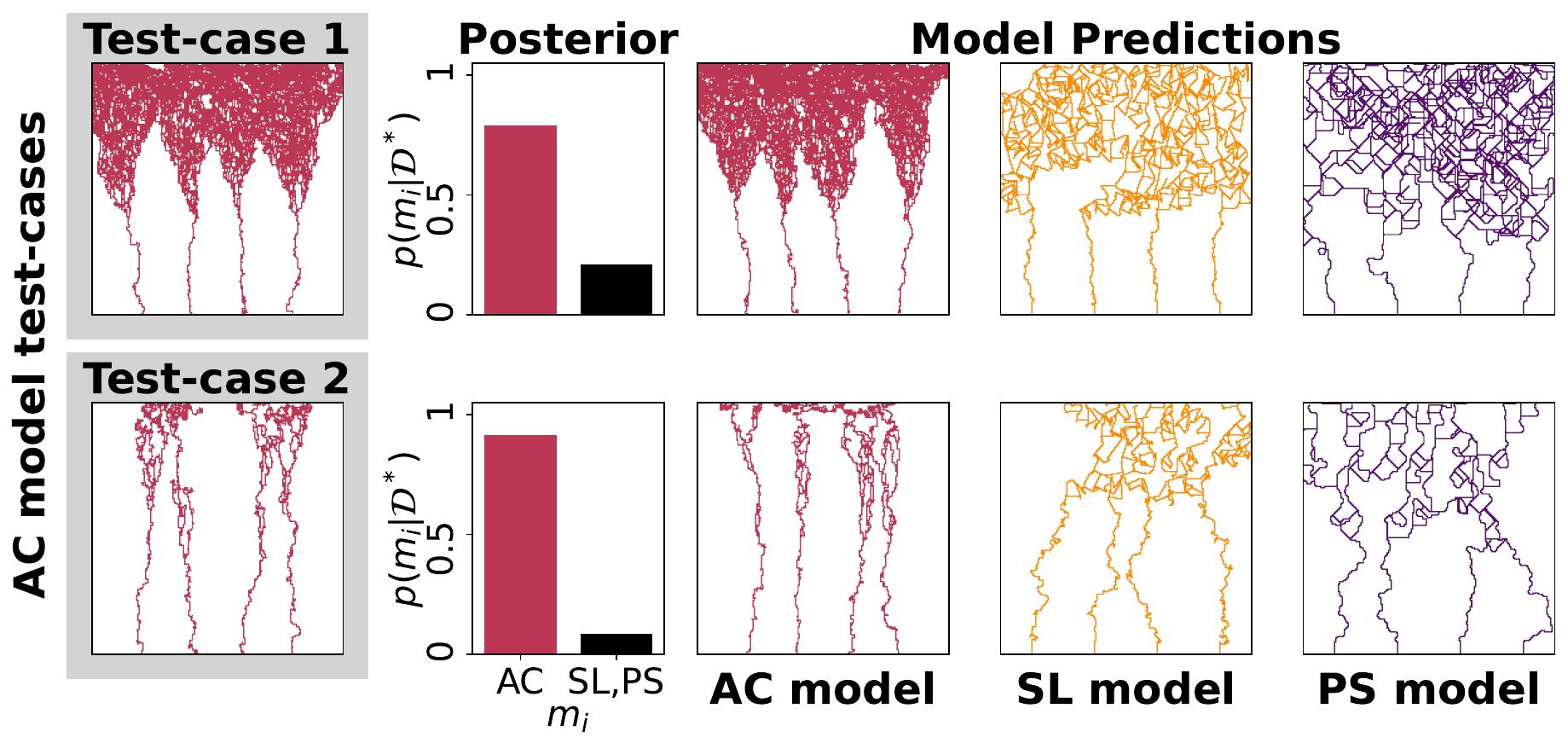}
\includegraphics[width=.48\textwidth]{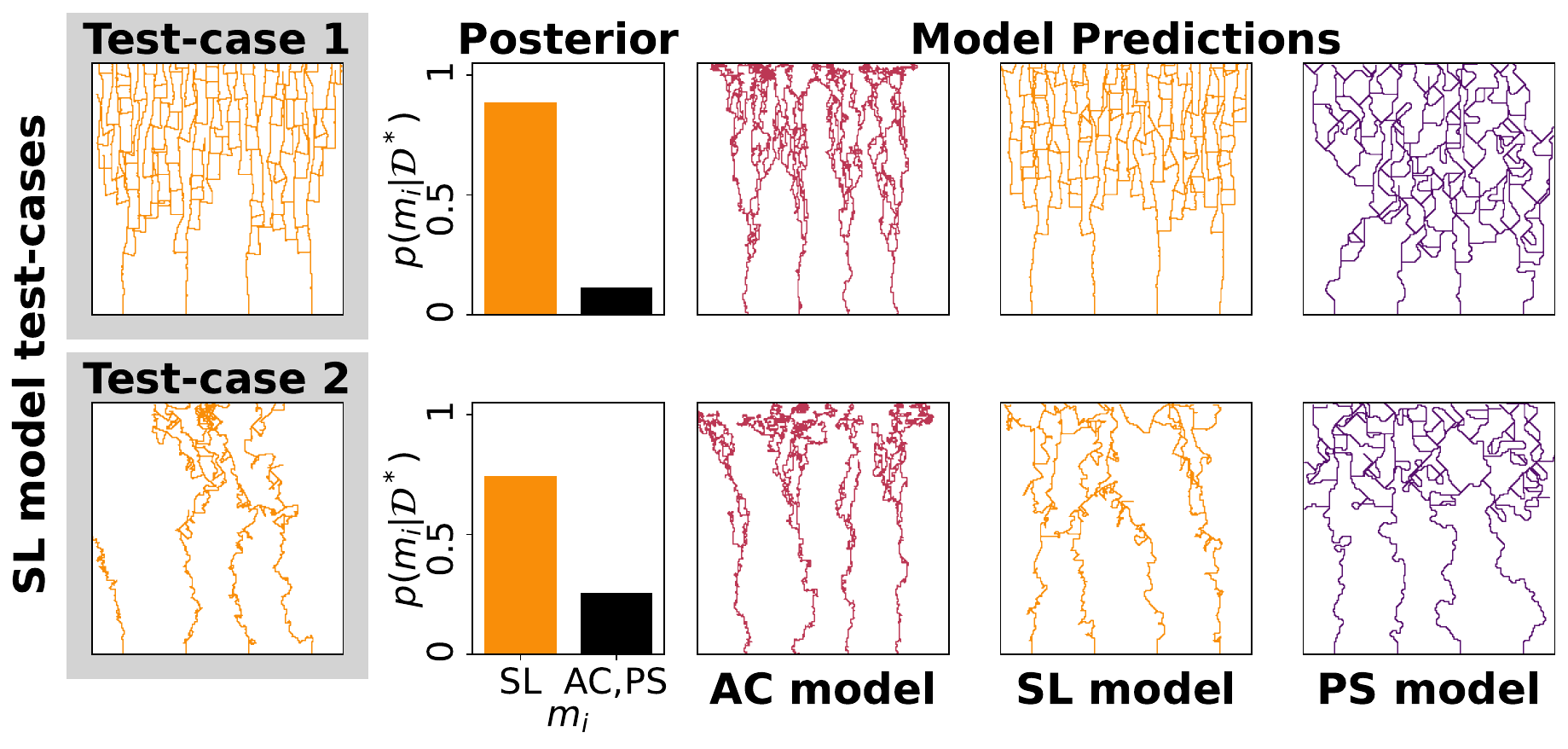}
\includegraphics[width=.48\textwidth]{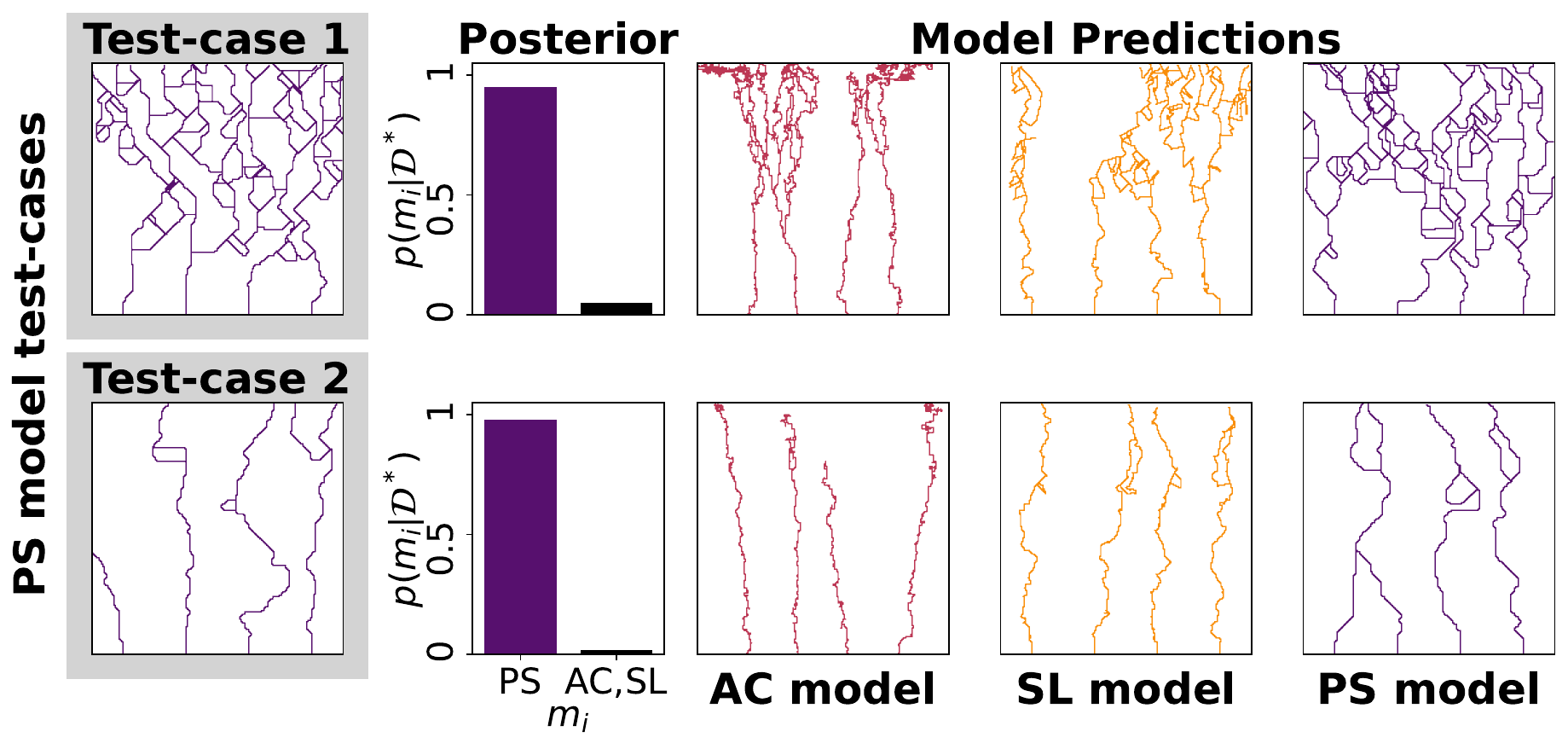}
\caption{We approximate the model posterior $p(m|\mathcal{D}^*)$ using the six test-cases from Figure \ref{fig:param_inference}, highlighting one example of $\mathcal{D}^*$.
For each test-case, we show one example of data simulated using an inferred parameter from each model's approximate parameter posterior.
Each `prediction' shows an example of that model's approximation of the true data generation process.}
\label{fig:model_selection}
\end{figure}

\subsection{ABC-SMC infers four parameters for each model and reproduces observed data}
We create two test-cases for each model by simulating data at known parameter values $10$ times.
Each test-case exhibits quantitatively different vascular networks that each model can produce. 
Following steps 1 and 2 of section \ref{sec:ourmethod}, we use ABC-SMC to infer four parameters for each model and show the resulting approximate posteriors in Figure \ref{fig:param_inference}.
In each test-case, the approximate posterior is unimodal and encompasses the true parameter, often close to its densest part.

\subsection{Random Forests correctly select models}
Using step 3 of section \ref{sec:ourmethod}, we approximate the model posterior for each of the six test-cases.
Figure \ref{fig:model_selection} shows the resulting approximate model posteriors, which identify the correct model with high probability in each test-case.
We simulate each model at a parameter value drawn from its approximate 
parameter posterior and show the resulting `prediction', which is a sample from the model's approximation of the true data generating process. 
Each model generates data that is visually similar to the observed data, however we can identify the true model in each test-case.

\section{Discussion}
Using Topological Data Analysis (TDA) and Approximate Bayesian Computation (ABC), we have developed a pipeline for parameter inference and model selection applicable to complex spatial models.
In previous work, TDA characterised the effect of two parameters in the AC model \citep{nardiniclustering} and was combined with ABC to infer them \citep{tabc}.
Here we extend this work by identifying a subset of informative summary statistics from multiple topological filtrations and use them to infer four parameters in three angiogenesis models using ABC-SMC.
We further show how RFs can be used with TDA to approximate model posteriors and compare candidate models.

While we validated our pipeline using synthetic data simulated from angiogenesis models, previous studies used \textit{in vitro} data to inform model rules and parameters \citep{mildeAngio, connorAngio, merksAngioModels}.
The present work therefore enhances previous model analysis, since ABC provides a statistical framework for learning parameters and evaluating models, and TDA provides a variety of filtrations and vectorisations which may be adapted to different spatial data.
In future work we will apply our pipeline to real experimental data, which will take the place of the synthetic test-cases in Figures \ref{fig:param_inference} and \ref{fig:model_selection} wherein the true parameter values and correct model will not be known.

In this study, we considered three models in which the paths traced by tip ECs form a static blood vessel network. 
Topological features, therefore, evolve monotonically in time in these models, and computing EPH at the final timestep was sufficient to infer parameters. 
In reality, sustained proliferation and vessel remodelling, where vasculature continually evolves after it is laid down, is characteristic of tumour-induced angiogenesis \citep{vascularremodeling}.
Indeed, vascular renormalisation, in which vessel-targetting agents prune small or inefficient blood vessels, is a theorised treatment strategy \citep{vascularnormalisation} aiming to temporarily enhance perfusion of the tumour to increase the effectiveness of radiotherapy \citep{jain_antiangio}.
Topological invariants that account for time-evolving data \citep{crokervineyardmulti}, directed flow networks \citep{grpph} and multiparameter filtrations \citep{mpp} could quantify such structural changes over time and be used to calibrate more sophisticated models.
Rather than fixing the duration of simulations in these cases, simulation time could be an extra parameter to be inferred by our pipeline.

Although we specialised our pipeline to discrete models of tumour-induced angiogenesis, its flexibility allows application to a range of spatio-temporal models.     
Any summary statistic which quantifies the desired properties of simulated data would be identified by the RF in step 1 if it captures the effect of changing model parameters. 
In future, we will use our pipeline to systematically compare continuum, cell-based, agent-based, and discrete models by their ability to reproduce observed data.

\ifthenelse{\boolean{arxiv_version}}{
    \begin{appendices}    
        \onecolumn
            \section{Angiogenesis Models}\label{SI:sec:models}
We use the Anderson-Chaplain (AC) model \citep{andersonChaplain}, the Stokes-Lauffenberger (SL) model \citep{stokesLauffenberger}, and the Plank-Sleeman (PS) model \citep{plank_sleeman} as our case-study for topological model selection. 
In this section, we present the models in dimensionless form, giving formulae and simulation details for each model.
We list model parameters in Table \ref{SI:tab:models_params} and provide illustrative schematics in Figure \ref{SI:fig:modelschematic}.

Each model simulates the movement of multiple tip endothelial cells (ECs) in a two-dimensional square domain $\mathcal{I} = \{(x, y): x, y \in [0, 1]\} \subset \mathbb{R}^2$.
The position of a tip EC at time $t$ is described by a variable $s^t = (s_x^t, s_y^t) \in \mathcal{I}$.
We initialise four tip ECs  along the lower edge of the domain (furthest from the tumour).
We create four variables $s$ such that ${s}^0 = (-1/8 + \upsilon/4, 0)$ for $\upsilon = 1, 2, 3, 4$.  
At discrete time intervals $\Delta t$, tip ECs migrate to a new position $s^{t + \Delta t} \in \mathcal{I}$, which is determined by a
set of model-specific movement rules.
In each model, we assume that a tumour is located along the domain's upper boundary ($y=1$), acting as a source of vascular endothelial growth factors (VEGF).
For simplicity, we prescribe the initial profile of VEGF $c(x, y, t=0)=y$ in which the VEGF concentration decreases linearly with distance from the tumour.
When a tip EC moves from $s^t$ and $s^{t + \Delta t}$, the line segment connecting subsequent positions is assumed to be occupied thereafter by immobile stalk EC (\textit{the snail-trail model} \citep{originalSnailTrail}).

\subsection{Anderson-Chaplain (AC) model}
The AC model employs an on-lattice biased random walk to simulate tip EC movement in response to local levels of VEGF and fibronectin.
A regular grid of $N\times N$ points spaced at intervals of $h$ is placed on the square domain $\mathcal{I}$ (we fix $N=200$ and $h=0.05$), and tip ECs move through the domain one grid space at a time.
The initial concentration of VEGF is as defined above and the initial concentration of fibronectin is  
$f(x, y,0) = 1-y$.
The time evolution of the VEGF and fibronectin concentrations, and the related probability that a tip EC moves left, right, down or up on the lattice, are derived from a system of Partial Differential Equations (PDEs).
\begin{align}   
\frac{\partial e}{\partial t} &= D \Delta e - \chi \nabla \cdot (e \nabla c) - \rho \nabla \cdot (e \nabla f) \label{eq:AC1}\\
\frac{\partial f}{\partial t} &= \beta e - \gamma e f \label{eq:AC2}\\
\frac{\partial c}{\partial t} &= -\eta e c\label{eq:AC3}
\end{align}
The PDEs \eqref{eq:AC1}--\eqref{eq:AC3}  describe how the spatial distribution of ECs, VEGF and fibronectin ($e, c, f: \mathcal{I} \times \mathcal{T}\to \mathbb{R}$) evolve over time.
The parameter $D$ determines the rate of EC random motility/diffusion, and parameters $\chi$ and $\rho$ give the strength of the ECs' chemotactic and haptotactic responses to spatial gradients of VEGF and fibronectin respectively.
ECs produce and degrade fibronectin at rates $\beta$ and $\gamma$, and consume VEGF at rate $\eta$. The PDEs are closed  by imposing no-flux boundary conditions along each side of the square domain $\mathcal{I}$.

To generate update rules for the concentration of VEGF and fibronectin at lattice points, as well as movement rules for the tip ECs, the PDE system is discretised using the Euler finite difference approximation.
Let $e^t_{l, m}, f^t_{l, m}, c^t_{l, m}$ be the values of $e$, $f$ and $c$ at lattice points $(lh, mh) \in \mathcal{I}$ at time $t$.
Discretising Equations \eqref{eq:AC1}--\eqref{eq:AC3} gives:
\begin{align}
    e^{t + \Delta t}_{l,m} &= e^{t}_{l,m} P_0 + e^{t}_{l+1,m} P_1 + e^{t}_{l-1,m} P_2 + e^{t}_{l,m+1} P_3 + e^{t}_{l,m-1} P_4,\label{eq:dAC1}\\
f^{t + \Delta t}_{l,m} &= f^{t}_{l,m} \left(1 - \Delta t \gamma e^{t}_{l,m}\right) + \Delta t \beta e^{t}_{l,m},\label{eq:dAC2}\\
c^{t + \Delta t}_{l,m} &= c^{t}_{l,m} \left(1 - \Delta t \eta e^{t}_{l,m}\right).\label{eq:dAC3}
\end{align}

On the lattice, the VEGF and fibronectin initial conditions become $c^0_{l,m} = m/N$ and  $f^0_{l,m} = 1- m/N$.
The discretisations \eqref{eq:dAC2}--\eqref{eq:dAC3} may then be used to update the VEGF and fibronectin concentrations $c^t_{l,m}$ and $f^t_{l,m}$ at grid locations $(l, m)$ in discrete time-steps of duration $\Delta t$ (using a discrete version of the no-flux boundary conditions).
Rather than using \eqref{eq:dAC2} to compute the concentration of ECs at lattice points, the factors $P_0, P_1, P_2, P_3$ and $P_4$, whose formulae are given in \eqref{eq:ACp0}--\eqref{eq:ACp4}, are used to determine the probability that an individual EC makes a move on the square lattice.

\begin{align}
    \begin{split}
    P_0 &= 1 - \frac{4\Delta t D}{h^2} - \frac{\Delta t \chi c^{t}_{l,m}}{h^2} \left(c^{t}_{l+1,m} + c^{t}_{l-1,m} - 4c^{t}_{l,m} + c^{t}_{l,m+1} + c^{t}_{l,m-1}\right) \\
&\ \ \ \ \ \  - \frac{\Delta t \rho}{h^2} \left(f^{t}_{l+1,m} + f^{t}_{l-1,m} - 4f^{t}_{l,m} + f^{t}_{l,m+1} + f^{t}_{l,m-1}\right)
\label{eq:ACp0} 
\end{split}
\\P_1 &= \frac{\Delta t D}{h^2} - \frac{\Delta t}{4h^2}\Big( \chi c^{t}_{l,m} (c^{t}_{l+1,m} - c^{t}_{l-1,m}) + \rho (f^{t}_{l+1,m} - f^{t}_{l-1,m})\Big)\label{eq:ACp1} \\
P_2 &= \frac{\Delta t D}{h^2} + \frac{\Delta t}{4h^2} \Big( \chi c^{t}_{l,m} (c^{t}_{l+1,m} - c^{t}_{l-1,m}) + \rho (f^{t}_{l+1,m} - f^{t}_{l-1,m}) \Big)\label{eq:ACp2}  \\
P_3 &= \frac{\Delta t D}{h^2} - \frac{\Delta t}{4h^2} \Big( \chi c^{t}_{l,m} (c^{t}_{l,m+1} - c^{t}_{l,m-1}) + \rho (f^{t}_{l,m+1} - f^{t}_{l,m-1})\Big)\label{eq:ACp3}  \\
P_4 &= \frac{\Delta t D}{h^2} + \frac{\Delta t}{4h^2} \Big( \chi c^{t}_{l,m} (c^{t}_{l,m+1} - c^{t}_{l,m-1}) + \rho (f^{t}_{l,m+1} - f^{t}_{l,m-1})\Big)\label{eq:ACp4} 
\end{align}
At each time step, each EC either remains at its location $(l, m)$, or moves to $(l-1, m)$, $(l+1, m)$, $(l, m-1)$ or $(l, m+1)$ according to the probabilities $\hat{P}_0, \hat{P}_1, \hat{P}_2, \hat{P}_3$ and $\hat{P}_4$, where $\hat{P}_j = P_j / (P_0+P_1+P_2+P_3+P_4)$ for $j = 0, 1, 2, 3, 4$.
To simulate this, a uniform random number $u\sim \mathcal{U}_{[0, 1]}$ is drawn for each tip EC at each time-step.
If $u \in [0, \hat{P}_0)$ then $s^{t+\Delta t} = s^t$, if $u \in [\hat{P}_0, \hat{P}_1)$, the EC moves left, and so on.
If this procedure prescribes a move outside of the domain $\mathcal{I}$, that tip EC terminates and is not considered for any further moves.

\subsection{Stokes-Lauffenberger (SL) model}
In the SL model \citep{stokesLauffenberger}, tip ECs can move in any direction (off-lattice). The movement of each tip EC is governed by \eqref{SI:eq:SLv}, a two-dimensional Stochastic Differential Equation (SDE) for EC velocity $\mathbf{V}(t)$.
At discrete time-steps, Equation \eqref{SI:eq:SLv} is solved using the Euler-Maruyama method and Equation \eqref{SI:eq:SLs} is integrated via the forward Euler method to give EC position.
\begin{align}
    \text{d}\mathbf{V}(t) &= - \beta \mathbf{V}(t)\,\text{d}t + \sqrt{\sigma} \,\text{d}\mathbf{W}(t) + \mathbf{\Psi}(t)\,\text{d}t \label{SI:eq:SLv}\\
    s(t) &= \int_0^t \mathbf{V}(\tau)\text{d}\tau
    \label{SI:eq:SLs}
\end{align}
In the SL model, the parameter $\beta$ models the cell's resistance to movement,
$\mathbf{W}(t)$ is a two-dimensional Wiener process (in which increments $\mathbf{W}(t + \Delta t) - \mathbf{W}(t)$ are independently and normally distributed), and the parameter $\sigma$ represents an EC's tendency to deviate from its current direction.
$\mathbf{\Psi}(t) = \kappa \nabla c \sin\abs{\frac{\psi}{2}}$ models movement due to chemotaxis, where $c$ is the VEGF concentration and $\psi$ is the angle that the current velocity $\mathbf{V}(t)$ makes with the direction of steepest increase in $c$.
Given $c(x, y) = y$, the direction of increasing VEGF concentration is always $(0, 1)^T$, which simplifies the chemotaxis term.
The parameter $\kappa$ measures the strength with which the EC velocity re-orients up spatial gradients of $c$ (towards the tumour).
ECs initialised along the bottom edge of the square domain are assigned a small initial velocity in the $y$-direction: $v^0 = (0, 0.5)^T$.
The discretisation \eqref{SI:eq:SLnumerical} of Equations \eqref{SI:eq:SLv} and \eqref{SI:eq:SLs} is then used to simulate EC velocity $v^t$ and position $s^t$ at each time-step.
\begin{equation}
         v^{t+\Delta t} = (1 - \beta \Delta t) v^t + \sigma \Delta t \varepsilon +  \kappa \Delta t 
         \sqrt{\frac{1 - v_y^t/\norm{v^t}}{2}}, \ \ 
     s^{t+\Delta t} = v^t + \Delta t  v^{t+\Delta t}
     \label{SI:eq:SLnumerical}
 \end{equation}
The quantity $\varepsilon \sim \mathcal{N}(\mathbf{0}, 1)$ is drawn from a two-dimensional normal distribution so that the random velocity vector has variance $\sigma$ in each direction.
Given the initial velocity, each tip EC in the SL model moves a distance $0.05$ (in the vertical direction) in the first time-step, which is the same as one grid-space in the AC model.
To ensure that ECs in the SL model move roughly this distance in every time-step (and the velocity does not grow exponentially), we fix $\beta = 0.8/\Delta  t$.

\subsection{Plank-Sleeman (PS) model}

The PS model \citep{plank_sleeman} assigns a constant speed $\hat{s}$ to each EC and varies the angle $\phi$ that a tip EC's velocity vector makes with the horizontal direction.
An EC's position $s^t = (s^t_x, s^t_y) \in \mathcal{I}$ is then modelled by the system of ordinary differential equations: 
\begin{equation}
    \frac{\text{d}s^t_x}{\text{d}t} = \hat{s}\cos{\phi}, \ \ \ \ \frac{\text{d}s^t_y}{\text{d}t} = \hat{s}\sin{\phi}.
    \label{eq:PSode}
\end{equation}
The movement angle is assumed to be independent of speed and position, and may be viewed as a random walk on the unit circle.
At each time-step, a tip EC may turn clockwise or counter-clockwise through a fixed angle $\hat{\phi}$ or it may continue in the same direction.
Given an initial movement angle $\phi_0$, the movement angle of each EC after $n$ time steps is determined by transition probabilities $\hat{\tau}^\pm_n$. 
The transition probabilities are derived from the mean turning rate $\mu(\phi)$, which is given by:

\begin{equation}
    \mu(\phi) = -d_c \abs{\nabla c} \sin(\phi - \phi_c).
    \label{SI:eq:PS_turning}
\end{equation}
In Equation \eqref{SI:eq:PS_turning}, the turning coefficient $d_c$ determines how often an EC angle re-orients its movement angle towards the direction of increasing VEGF concentration 
$\phi_c = \pi/2$.
Let $\hat{\tau}^\pm_n$ denote the probability that an EC rotates through a fixed angle of $\pm \hat{\phi}$ on the $n$-th time-step.
It can be shown \citep{plank_sleeman} that, if the mean turning rate is defined by \eqref{SI:eq:PS_turning}, then $\hat{\tau}^\pm_n$ are given by:
\begin{equation}
\hat{\tau}^{\pm}_n = 2\lambda\frac{ \tau \left( \left(n \pm \frac{1}{2}\right)\hat{\phi} \right)}{\tau \left( \left(n + \frac{1}{2}\right)\hat{\phi} \right) + \tau \left( \left(n - \frac{1}{2}\right)\hat{\phi} \right)}\text{,    where }
\tau(\phi) = \exp \left(\frac{d_c}{D_r} \cos(\phi - \phi_c) 
\right)
\text{    and   } \lambda = D_r/\hat{\phi}^2.
\end{equation}
Choosing a random number $u\sim \mathcal{U}_{[0, 1]}$ from the standard uniform distribution, an EC turns anticlockwise through an angle of $\hat{\phi}$ if $u \in [0, \hat{\tau}^+_n \Delta t)$, clockwise through an angle of $\hat{\phi}$ if $u \in [\hat{\tau}^+_n \Delta t, 2\lambda \Delta t)$, and continues in its current direction otherwise.
Using this rule to generate movement angles $\phi_t$ at time steps $t \in \mathcal{T}$, the position $s^t = (s^t_x, s^t_y) \in [0, 1]^2$ of an individual EC is determined by solving the ODEs \eqref{eq:PSode} using the forward Euler method:
\begin{equation}
    s^{t+\Delta t}_x = s^t_x +\hat{s} \Delta t  \cos{\phi_t}, \ \ \ \
     s^{t+\Delta t}_y = s^t_y + \hat{s} \Delta t \sin{\phi_t}.
\end{equation}
Choosing a speed of $\hat{s} = 0.05$ ensures that, in the PS model, ECs move the same distance during each time-step as ECs in the AC model, and a similar distance as ECs in the SL model.

\subsection{Anastomosis, branching and termination rules for all models}
The previous sections describe how each model determines an EC's new position $s^{t + \Delta t}$ from its current position $s^t$.
The rules for EC branching, anastomosis and termination are the same for all three models.

An active EC at location $s^t$ may bifurcate into two separate EC.
After branching, the original tip EC continues to move as instructed by its model, and a new variable $s^t$ is created at the branch point to represent a new tip EC, which thereafter moves independently.
We denote by $a_\text{br}$ the minimum age that a EC must reach before it can be considered for a bifurcation.
A second branching parameter, $c_\text{br}$, defines the minimum concentration of VEGF that must be present at $s^t$ in order for the tip EC to bifurcate.
In all models, EC bifurcate into two separate EC as soon as both the minimum age for branching and the VEGF threshold for branching have been exceeded.
Since we initiate multiple ECs in the simulation domain, it is possible that a move may result in a tip EC crossing the path of an existing stalk EC (or colliding with another tip EC).
In such cases, all three models assume an anastomosis event occurs.
In the AC model, if a move requires an EC to move into a grid position that is already occupied, the EC does not make this move; it is terminated and not considered for any subsequent moves.
In the SL and PS models, if a move requires an EC to cross an existing EC path, the EC  terminates at the intersection of the proposed move and the existing path.
If a model's movement rule specifies a new position $s^{t+\Delta t}$ which is outside the simulation domain $\mathcal{I}$, that tip EC terminates and is not considered for further movement.

Figure \ref{SI:fig:modelschematic} gives an illustration of EC movement rules, including examples of anastomosis, which can lead to loops in the simulated network.

\newpage

\subsection{Model simulation, schematic and parameters}
In each model, we use a timestep of $\Delta t = 0.01$, and simulate EC movement over the (dimensionless) time interval $\mathcal{T} = [0, t_\text{final}]$.
We wish to fix $t_\text{final}$ to allow a fair comparison between models, and large enough to ensure we observe enough EC movement to infer model parameters.
ECs move approximately 0.005 spatial units during each timestep in each model--this is the exactly the grid spacing in the AC model, and we scaled the SL and PS models so that ECs move approximately this distance in each timestep too.
If all ECs moved directly upwards in each timestep, they would reach the tumour at time $t=2$.
We choose $t_\textnormal{final}=4$, thus simulating each model for $400$ timesteps, which will allow most meandering or bifurcated ECs to terminate by either reaching the tumour, anastomosing, or moving outside of the simulation domain.

We fix all but four parameter values in each angiogenesis model--see Table \ref{SI:tab:models_params} for a list of the parameters that vary and their ranges.
See also the schematic at the top of Figure \ref{SI:fig:modelschematic} for an illustration of model movement rules.

\begin{figure}[h]
    \centering
    \includegraphics[width=\linewidth]{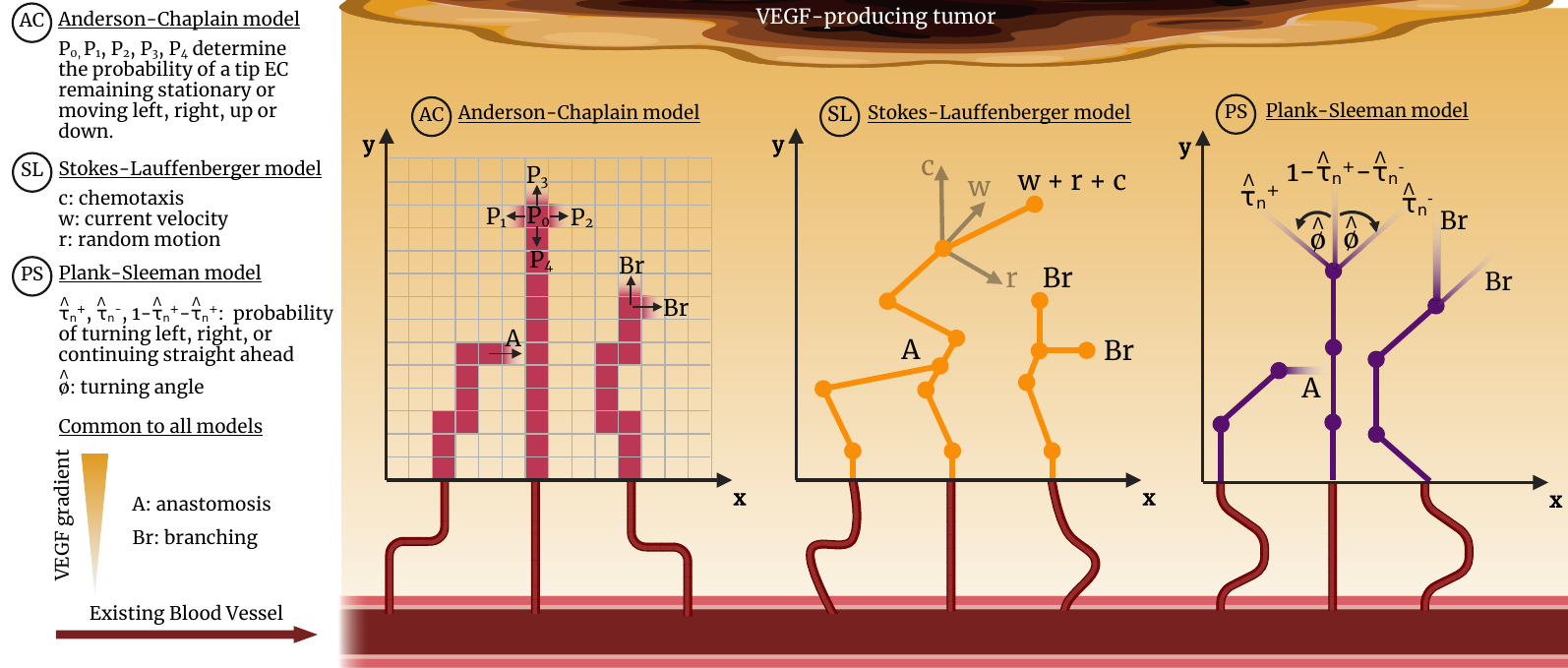}
    \caption{Schematic showing tip EC movement rules in the Anderson-Chaplain (AC), Stokes-Lauffenberger (SL) and Plank-Sleeman (PS) models.
In the AC model, tip ECs move on a grid according to probabilities $\hat{P}_j = P_j / (P_0 + P_1 + P_2 + P_3 + P_4)$ for $j=0, 1, 2, 3, 4$. Higher values of the chemotaxis and haptotaxis parameters induce a bias into those probabilities  which specify a movement towards increasing concentrations of VEGF and fibronectin respectively.
The SL model updates the velocity $v_i^t$ of the EC at location $s^t$ as a weighted sum of the current velocity $w$, randomness $r$ and chemotaxis $c$. 
Randomness and chemotaxis and parameters regulate the weight of the corresponding terms when updating the velocity.
The PS model rotates the movement angle $\phi$ between an EC's velocity vector and the horizontal by $\hat{\phi}$ with transition probabilities $\hat{\tau}_n^+$ and $\hat{\tau}_n^-$ and
ECs move a fixed distance $\hat{s}$ in the new direction at each time-step.
A turning rate parameter regulates how often the EC's angle of movement updates, and a turning bias parameter changes how likely such an update is to favor the direction of increasing VEGF concentration.
All models use the same rules for branching--a tip EC at location $s^t$ bifurcates into two tip ECs that move independently when $t$ is greater than the minimum age for branching parameter $a_\text{br}$ and the concentration of VEGF at $s^t$ is greater than the VEGF threshold for branching parameter $c_\text{br}$.
If a movement rule would cause an EC to move into a location already occupied by an EC, that EC instead anastomoses and is considered for no further movement.
Created in \url{https://BioRender.com}}
\label{SI:fig:modelschematic}
    \end{figure}\clearpage
    \begin{table}[]
        \begin{tabular}{|p{24mm}p{24mm}p{24mm}||p{24mm}p{24mm}p{24mm}||p{24mm}p{24mm}p{25mm}|}
\hline
\multicolumn{3}{|c||}{\textbf{Anderson-Chaplain (AC) model}} & \multicolumn{3}{c||}{\textbf{Stokes-Lauffenberger (SL) Model}} & \multicolumn{3}{c|}{\textbf{Plank-Sleeman (PS) Model}} \\ \hline
\multicolumn{3}{|c||}{parameters} & \multicolumn{3}{c||}{parameters} & \multicolumn{3}{c|}{parameters} \\ \hline
\multicolumn{1}{|l|}{\begin{tabular}[c]{@{}l@{}}Name and\\ Symbol\end{tabular}} & \multicolumn{1}{l|}{Range} & Notes & \multicolumn{1}{l|}{\begin{tabular}[c]{@{}l@{}}Name and\\ Symbol\end{tabular}} & \multicolumn{1}{l|}{Range} & Notes & \multicolumn{1}{l|}{\begin{tabular}[c]{@{}l@{}}Name and\\ Symbol\end{tabular}} & \multicolumn{1}{l|}{Range} & Notes \\ \hline
\multicolumn{1}{|l|}{\begin{tabular}[c]{@{}l@{}}Chemotaxis\\ $\chi$\end{tabular}} & \multicolumn{1}{l|}{$[0, 0.5]$} & Taken from \citep{nardiniclustering} & \multicolumn{1}{l|}{\begin{tabular}[c]{@{}l@{}}Chemotaxis\\ $\kappa$\end{tabular}} & \multicolumn{1}{l|}{$[0, 100]$} & \begin{tabular}[c]{@{}l@{}}The chemotaxis \\ term in \eqref{SI:eq:SLnumerical} has \\ the same order of\\ magnitude as the \\ current velocity\end{tabular} & \multicolumn{1}{l|}{\begin{tabular}[c]{@{}l@{}}Turning \\ bias\\ $d_c$\end{tabular}} & \multicolumn{1}{l|}{$[30, 150]$} & \begin{tabular}[c]{@{}l@{}}30 gives no turning\\ bias, 150 almost\\ always favours\\ turning towards\\ the VEGF source\end{tabular} \\ \hline
\multicolumn{1}{|l|}{\begin{tabular}[c]{@{}l@{}}Haptotaxis\\ $\rho$\end{tabular}} & \multicolumn{1}{l|}{$[0, 0.5]$} & Taken from \citep{nardiniclustering} & \multicolumn{1}{l|}{\begin{tabular}[c]{@{}l@{}}Randomness\\ $\sigma$\end{tabular}} & \multicolumn{1}{l|}{$[0, 100]$} & \begin{tabular}[c]{@{}l@{}}The randomness \\ term in \eqref{SI:eq:SLnumerical} has \\ the same order of\\ magnitude as the \\ current velocity\end{tabular} & \multicolumn{1}{l|}{\begin{tabular}[c]{@{}l@{}}Turning \\ rate\\ $D_r$\end{tabular}} & \multicolumn{1}{l|}{$[0, 30]$} & \begin{tabular}[c]{@{}l@{}}A turning rate of \\ 30 means EC \\ change direction\\ every other\\ time-step\end{tabular} \\ \hline
\multicolumn{1}{|l|}{\begin{tabular}[c]{@{}l@{}}Minimum\\ age for\\ branching\\ $a_\text{br}$\end{tabular}} & \multicolumn{1}{l|}{$[0, 1]$} & \begin{tabular}[c]{@{}l@{}}ECs can \\ bifurcate after\\ between 0 and\\ 100 time-steps.\end{tabular} & \multicolumn{1}{l|}{\begin{tabular}[c]{@{}l@{}}Minimum\\ age for\\ branching\\ $a_\text{br}$\end{tabular}} & \multicolumn{1}{l|}{$[0, 1]$} & \begin{tabular}[c]{@{}l@{}}ECs can \\ bifurcate after\\ between 0 and\\ 100 time-steps.\end{tabular} & \multicolumn{1}{l|}{\begin{tabular}[c]{@{}l@{}}Minimum\\ age for\\ branching\\ $a_\text{br}$\end{tabular}} & \multicolumn{1}{l|}{$[0, 1]$} & \begin{tabular}[c]{@{}l@{}}ECs can \\ bifurcate after\\ between 0 and\\ 100 time-steps.\end{tabular} \\ \hline
\multicolumn{1}{|l|}{\begin{tabular}[c]{@{}l@{}}VEGF \\ threshold\\ for\\ branching\\ $c_\text{br}$\end{tabular}} & \multicolumn{1}{l|}{$[0, 1]$} & \begin{tabular}[c]{@{}l@{}}The concentration \\ of VEGF varies in\\ $\mathcal{I}$ from 0 when \\ $y=0$ to 1 when\\ $y=1$.\end{tabular} & \multicolumn{1}{l|}{\begin{tabular}[c]{@{}l@{}}VEGF \\ threshold\\ for\\ branching\\ $c_\text{br}$\end{tabular}} & \multicolumn{1}{l|}{$[0, 1]$} & \begin{tabular}[c]{@{}l@{}}The concentration \\ of VEGF varies in\\ $\mathcal{I}$ from 0 when \\ $y=0$ to 1 when\\ $y=1$.\end{tabular} & \multicolumn{1}{l|}{\begin{tabular}[c]{@{}l@{}}VEGF \\ threshold\\ for\\ branching\\ $c_\text{br}$\end{tabular}} & \multicolumn{1}{l|}{$[0, 1]$} & \begin{tabular}[c]{@{}l@{}}The concentration \\ of VEGF varies in\\ $\mathcal{I}$ from 0 when \\ $y=0$ to 1 when\\ $y=1$.\end{tabular} \\ \hline

\end{tabular}   
\caption{\rmfamily The parameters inferred in each model and their values/ranges.}
\label{SI:tab:models_params}
    \end{table}
\clearpage
    
\section{Computation of Summary Statistics}\label{SI:sec:ss}
We convert data simulated from the three angiogenesis models (Anderson-Chaplain (AC), Stokes-Lauffenberger (SL), Plank-Sleeman (PS)) into a common format and then compute summary statistics.

Although simulations describe the movement of ECs in the domain $\mathcal{I}$ throughout the time interval $[0, t_\text{final}]$, EC trails do not remodel after they have been laid down in the models we study.
The value $t_\text{final} = 4$ we choose allows the majority of ECs to reach the tumour within the simulation time (see the previous section for details).
Therefore, the snapshot of the domain at the final timestep contains most of the information provided by the full simulation.
Hence, and in order to simplify calculations, we consider only this final timestep snapshot when computing summary statistics.

The common format retains the spatial structure of simulated networks while discretising them to allow the computation of spatially-averaged and topological summary statistics.
Considering a simulated network at its final timestep, we overlay a regular square grid of $K=200$ points spaced in intervals of $h=0.05$ in co-ordinate directions.
We then say an \textit{angiogenesis dataset} $\mathcal{D}$ is the point-cloud consisting of the $(x, y)$ locations of ECs within this discretised image.
The AC model simulates data on such a grid already, and we convert data from the SL and PS models into the common format by populating those grid squares which intersect the (off-lattice) paths traced by tip ECs.
This common format will also be applied to observed data and it ensures that summary statistics give a fair comparison between data simulated by each model. 
\subsection{Spatially-Averaged Summary Statistics} 
An angiogenesis dataset $\mathcal{D}$ is a point-cloud consisting of $(x, y)$ grid locations which contain simulated EC at the final simulation timestep.
We compute the mean, standard deviation, minimum, maximum, range, interquartile range, and the 10th, 25th, 75th and 90th percentiles of the $x$ and $y$-coordinates of points in $\mathcal{D}$.
Concatenating these values gives a list of $20$ spatially-averaged summary statistics, $10$ in the horizontal ($x$) coordinate direction and $10$ in the vertical ($y$) coordinate direction.

\subsection{Extended Persistent Homology}\label{SI:subsec:extpers}
We detect and quantify topological structure in angiogenesis datasets using extended persistent homology (EPH).
Here, we give details of how we compute and vectorise EPH, outlining how it arises from persistent homology (PH) and explaining our reasons for using EPH instead of PH.

To compute PH, spatial data is first converted to a nested sequence of simplicial complexes $\{\Sigma_k\}_{k=0}^K$ known as a filtration.
A simplicial complex is a collection of vertices ($0$-simplices), where subsets of vertices can be connected by edges ($1$-simplices), triangles ($2$-simplices), tetrahedra ($3$-simplices) and their higher dimensional analogues.
A face of a simplex is defined as a subset of its vertices along with the simplices that are connected to them.
Simplicial complexes satisfy the property that a face of any simplex, or the intersection of two simplices, is also a simplex in the complex.

The dimension-$p$ PH of the filtration is then the sequence \eqref{SI:eq:persmod} of $\mathbb{F}$-vector spaces $H_p(\Sigma_k)$  (we fix $\mathbb{F}=\mathbb{Z}/2\mathbb{Z}$) together with maps induced by the inclusion maps between the simplicial complexes.
The basis elements of $H_p(\Sigma_k)$ correspond to $p$-dimensional holes in $\Sigma_k$ \citep{carlssonTDA}, also referred to as topological features of the simplicial complex.
Since each $\Sigma_k \subset \Sigma_{k+1}$, inclusion maps $\iota_k:\Sigma_k \to \Sigma_{k+1}$ map simplices in $\Sigma_k$ to their counterparts in the larger $\Sigma_{k+1}$.
The inclusion maps $\iota_k$ then induce linear maps $\iota_k^*: H_p(\Sigma_k) \to H_p(\Sigma_{k+1})$ in sequence \eqref{SI:eq:persmod} (by functoriality) which allow topological features to be tracked through the filtration.
\begin{align}
    H_p(\Sigma_{0}) 
    \xrightarrow{\iota_0^*} H_p(\Sigma_{1}) \xrightarrow{\iota_1^*} 
    \dots \xrightarrow{\iota^*_{k-1}} H_p(\Sigma_k) \xrightarrow{\iota^*_k}
    \dots \xrightarrow{\iota_{K-2}^*} H_p(\Sigma_{K-1})
    \xrightarrow{\iota_{K-1}^*} H_p(\Sigma_K)
    \label{SI:eq:persmod}
\end{align} 
The Structure Theorem of PH \citep{structurePH} states that sequence \eqref{SI:eq:persmod} uniquely decomposes into a direct sum of interval modules $I_{b, d}$, which are sequences of $\mathbb{F}$-vector spaces \eqref{SI:eq:intervalmod}, where $\eta_k^*$ are identity maps when $b \leq k < d$ and zero maps otherwise.
\begin{align}
 0
    \xrightarrow{\eta_0^*}0  \dots 0
    \xrightarrow{\eta_{b-1}^*} 
    \mathbb{F} \xrightarrow{\eta_{b}^*}
\mathbb{F}
\dots \mathbb{F}
    \xrightarrow{\eta_{d-1}^*} 
    0  \dots 0\xrightarrow{\eta_{K-1}^*} 0\label{SI:eq:intervalmod}
\end{align}
Persistence pairs $(b, d)$ that define interval modules \eqref{SI:eq:intervalmod} correspond to topological features in the filtration.
The birth $b$ is the first index $k$ for which the corresponding $p$-dimensional hole appears in the filtration (and the first index $k$ in \eqref{SI:eq:intervalmod} for which the corresponding topological feature is in the image of $\eta_k^*$ but not in the image of $\eta_{k-1}^*$).
The death $d$ is the filtration index $k$ where the $p$-dimensional hole is filled in by additional simplices (and the first index $k$ in \eqref{SI:eq:intervalmod} for which the corresponding topological feature is mapped to $0$ by $\eta_{k-1}^*$).
The persistence $d-b$ measures how long the corresponding topological feature persists in the filtration.
Persistence pairs may be plotted as points in birth-death coordinates in a persistence diagram (PD).
PDs are stable \citep{PHstable} to small perturbations in the underlying point-cloud, making them useful topological summaries of spatial data.

\cite{nardiniclustering} applied PH to study angiogenesis datasets $\mathcal{D}$ by constructing a sweeping-plane filtration as follows. 
First, $\mathcal{D}$ is converted to a simplicial complex $\Sigma$.
Each point $(x, y)$ in the angiogenesis dataset is represented in $\Sigma$ by a vertex ($0$-simplex).
If two vertices represent two grid locations in $\mathcal{D}$ which are adjacent (in the Moore neighbourhood), they are connected with edges ($1$-simplices).
Collections of three edges are connected with triangle ($2$-simplex) if each pair of edges shares a vertex.
A function $f: \Sigma \to \mathbb{R}$ is then defined such if a vertex $u$ represents a point $(x, y) \in \mathcal{D}$, $f(u) = y$. 
The value of $f$ on other simplices within $\Sigma$ is then simply the maximum of the values of $f$ on their vertices. 
The sweeping-plane filtration is then defined as the sublevel sets $\Sigma_k = f^{-1}(-\infty, k/K]$, where $k=0, 1, \dots, K$ and $K$ is the resolution of the image (e.g. $K=200)$.

Using this filtration, persistence pairs quantify the location of components and loops (measured in the vertical ($y$) coordinate direction).
However, some persistence pairs computed using this method have infinite persistence, since topological features often persist throughout the entirety of the sequence \eqref{SI:eq:persmod} (and $d=\infty$). For example, all loops persistent infinitely in this filtration.
In these cases, the size of the $p$-dimensional holes is not recovered by the persistence pair $(b, d)$.
Here, we wish to quantify both the size and location of topological features in angiogenesis datasets, and we wish to precisely measure branch points, loops, anastomoses and components, which PH with the above sweeping-plane filtration is not able to do.
We therefore turn to extended persistent homology (EPH).

To compute EPH, one defines $\Sigma^k$ as the simplicial complex containing those simplices in $\Sigma_K$ but not $\Sigma_k$.
With the sweeping-plane filtration above, $\Sigma^k$ are superlevel sets $f^{-1}[k/K, \infty)$. 
The relative homology $H_p(\Sigma_K, \Sigma^k)$ then quantifies topological features in the quotient complex $\Sigma_K/\Sigma^k$.
Since $\Sigma^{k+1} \supset \Sigma^{k}$, a quotient map $q_k: \Sigma^{k+1} \to \Sigma^{k}$ may be defined for $k = K-1, \dots, 0$.
$q_k$ maps all simplices in $\Sigma^{k+1}$ but not $\Sigma^{k}$ to a single point and is the identity map when restricted to $\Sigma^{k}$.
The quotient maps $q_k$ induce linear maps in sequence \eqref{SI:eq:extpersmod} which track topological features through the relative homology groups.
\begin{align}
    H_p(\Sigma_K, \Sigma^{K}) 
    \xrightarrow{q_{K-1}^*} H_p(\Sigma_K, \Sigma^{K-1}) \xrightarrow{q_{K-2}^*} 
    \dots \xrightarrow{q^*_{k}} H_p(\Sigma_K, \Sigma^k) \xrightarrow{q^*_{k-1}}
    \dots \xrightarrow{q_{1}^*} H_p(\Sigma_K, \Sigma^{1})
    \xrightarrow{q_{0}^*} H_p(\Sigma_K, \Sigma^0)
    \label{SI:eq:extpersmod}
\end{align}
Since $H_p(\Sigma_K) = H_p(\Sigma_K, \Sigma^K)$, the sequences \eqref{SI:eq:persmod} and \eqref{SI:eq:extpersmod} may be concatenated to give a single sequence (of length $2k + 1$), which is known as the dimension-$p$ EPH of the filtration.
The structure theorem applies to this sequence, meaning that topological features may be extracted as persistence pairs $(b, d)$.
Intuitively, the sweeping-plane filtration scans through $\Sigma$ from bottom to top and the vector spaces \eqref{SI:eq:persmod} detect topological features which are found below $k/K$.
The complexes $\Sigma_K/\Sigma^k$ then scan back down from top to bottom, collapsing all simplices above $k/K$ to a single point, with the sequence \eqref{SI:eq:extpersmod} detecting topological features in the resulting complexes.
In particular, all simplices merge in $\Sigma_K / \Sigma^0$, and no persistence pairs extracted from the combined sequence \eqref{SI:eq:persmod}--\eqref{SI:eq:extpersmod} have $d=\infty$.

All features computed through EPH are born and die at some point in the concatenated sequence \eqref{SI:eq:persmod}--\eqref{SI:eq:extpersmod}.
Features may be classified into four types depending depending on which part of this sequence they are born and die in.
If a feature is born and dies within sequence \eqref{SI:eq:persmod}, it is called ordinary.
If a feature is born and dies within in sequence \eqref{SI:eq:extpersmod}, it is called relative.
If a feature is born in sequence \eqref{SI:eq:persmod} and dies in sequence \eqref{SI:eq:extpersmod} it is called extended.
Such persistence pairs are called extended+ if $b>d$ and extended- if $b<d$.

We use two functions $f$ to create two sweeping-plane filtrations from an angiogenesis dataset $\mathcal{D}$.
After converting $\mathcal{D}$ into a simplicial complex $\Sigma$ as above, we define $f_\text{horizontal}, f_\text{vertical}: \Sigma \to \mathbb{R}$ on vertices $u$ representing points $(x, y) \in \mathcal{D}$ by $f_\text{horizontal}(u) = x$, $f_\text{vertical}(u) = y$, and on other simplices as before.
Using these filtrations to compute EPH, persistence pairs resulting from the sequences \eqref{SI:eq:persmod}--\eqref{SI:eq:extpersmod} then represent the size and location of topological features measured in the horizontal ($x$) and vertical ($y$) coordinate directions.
Persistence pairs may be plotted on an extended persistence diagram (EPD).
Figure \ref{SI:fig:topfeatures} contains an example computation of EPDs using the horizontal and vertical sweeping-plane filtration.
Each EPD contains one feature of each type (ordinary, relative, extended+, and extended-).
This Figure is an expanded version of Figure 2 of the main text, and it also shows the extra information gained by computing EPH over PH.

\subsection{Topological Summary Statistics}\label{SI:subsec:vectorisations}
We use Persistence Images and persistence statistics to transform EPDs into fixed-length vectors amenable to further analysis.

Persistence Images are stable vector representations of persistence diagrams \citep{persistenceimage}, and they were used by \cite{tabc} as summary statistics to infer two parameters in the AC model. 
To obtain Persistence Images, persistence pairs $(b, d)$ are first transformed to the modified pair $(b, d-b)$. 
A persistence surface is then the function $\Xi : \mathbb{R}^2 \to \mathbb{R}$ \eqref{SI:eq:psurface}, where the sum is taken over all modified pairs.
\begin{equation}
 \Xi(x, y) = \sum_{(b, d-b)} w(b, d-b)\Phi_{(b, d-b)}(x,y).
 \label{SI:eq:psurface}
\end{equation}
Here, $w$ is a weighting function on the modified persistence pairs, and we use the standard choice $w(b, d-b) = d-b /\max{(d-b)}$, which divides the persistence by the maximum persistence across all persistence pairs.
The function $\Phi_{(b, d-b)}$ must be a differentiable probability distribution with mean $(b, d-b)$, and we take the standard Gaussian distribution with variance $0.1$.
A Persistence Image is then a collection of integrals \eqref{SI:eq:pimage} of $\Xi$ over discretised regions $\mathcal{R} \subset \mathbb{R}^2$:

\begin{equation}
    \Gamma(\Xi)_\mathcal{R} = \int\int_\mathcal{R} \Xi \text{d}y \text{d}x.
    \label{SI:eq:pimage}
\end{equation}

The regions $\mathcal{R}$ must be chosen so that the integrals reflect the distribution of persistence pairs used to construct $\Xi$.
The persistence pairs we compute from angiogenesis datasets with sweeping-plane filtrations have birth and death values between $k=0$ and $k=200$.
We therefore take regions $R_{i, j} = [20i, 20(i+1)] \times [20j, 20(j+1)]$ for $i, j = 0, 1, \dots, 9$ to generate a total of $100$ integrals which constitute a Persistence Image. 
For each sweeping-plane filtration (horizontal and vertical), we compute four persistence images--one for each of the four types of extended persistence pair described in section \ref{SI:subsec:extpers} .
Therefore, Persistence Images give a total of $800$ topological summary statistics for each angiogenesis dataset.

Persistence statistics are a way to summarise a collection of persistence pairs, which have been found to perform well on well-known classification tasks \citep{vectorisationSurvey}.
To obtain persistence statistics, one considers the births $b$, deaths $d$, persistences $d-b$ and midpoints $(b+d)/2$ of persistence pairs $(b, d)$, and computes the mean, standard deviation, median, interquartile range, full range, and the 10th, 25th, 75th and 90th percentiles of each of these quantities.
Within each EPD, we consider each type of extended persistence pair (ordinary, relative, extended+, and extended-) separately, since each quantifies a different spatial feature within angiogenesis data (see Figure \ref{SI:fig:topfeatures} for examples).
We therefore compute a total of $36$ persistence statistics for each of the four types of persistence pairs described in section \ref{SI:subsec:extpers} for each sweeping-plane filtration (horizontal and vertical)
Therefore, persistence statistics give a total of $36\times4\times2 = 288$ topological summary statistics for each angiogenesis dataset.

\subsection{Example Computation}
We compute spatially-averaged and topological summary statistics from angiogenesis datasets and concatenate these into a long-list of summary statistics.
Each angiogenesis dataset yields $20$ spatially-averaged summary statistics and $800+288$ topological summary statistics, which we combine into a vector of length $1108$.
An example of the computation of spatially-averaged summary statistics and EPDs for a simple angiogenesis dataset is given in Figure \ref{SI:fig:topfeatures}.\\

\subsection{Distribution of Important Summary Statistics}
In Figure 1 of the main text, we report the number of each type of summary statistics which were selected as informative for each model by step 1 in section 4 of the main text. 
Figure \ref{SI:fig:top_ss} shows the breakdown of the top 100 summary statistics for each parameter in each model by type (spatially-averaged or topological), direction (vertical or horizontal), as well as vectorisation (Persistence Images or persistence statistics).
We found that a mixture of Persistence Images and persistence statistics are informative for inferring parameter values.

No spatially-averaged summary statistics computed in the horizontal ($x$) direction are selected by step 1 of our pipeline for any parameter in any model.
The $x$ co-ordinates of EC locations are clustered around the four initialisation points on the horizontal axis in simulations.
Some parameter values cause ECs to remain tightly clustered around these four points, and others cause them to spread out.
However, horizontal spatially-averaged statistics are too coarse to distinguish these scenarios, which may explain why they are not useful in parameter inference.

\begin{figure}[!hb]
    \centering
\includegraphics[width=.825\linewidth]{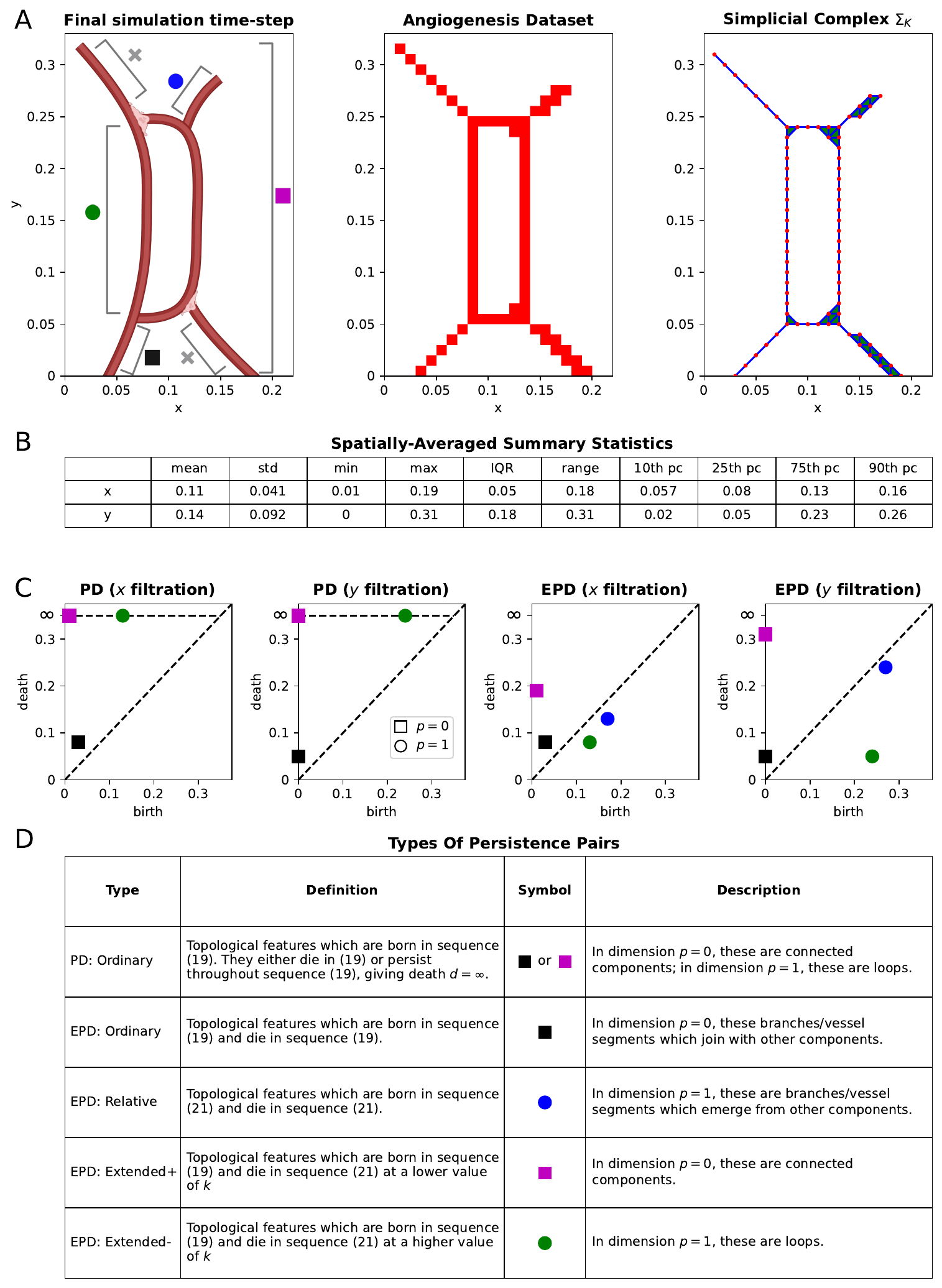}
    \caption{
    \textbf{An example model simulation and computation of spatially-averaged summary statistics, persistence diagrams (PDs) and extended persistence diagrams (EPDs)}.
    A) Data pre-processing.
    We display the final time-step of a simple simulation in which two ECs migrate from the bottom of the simulation domain, branch, anastomose and form a loop in their path upward towards the tumour.
    We convert the final simulation time-step into an angiogenesis dataset by overlaying a grid of $200\times 200$ pixels onto the simulation domain and colouring pixels which correspond to the location of simulated ECs.
    Simplicial complexes $\Sigma_k$ contain a vertex for each pixel location containing an EC whose $x$ co-ordinate (or $y$ coordinate) is less than or equal to $k/200$--we plot $\Sigma_K$ where $K=200$.
    B) Spatially-averaged summary statistics.
    We compute 10 simple statistics on the $x$ and $y$ co-ordinates of pixel locations which contain ECs. 
    C)-D) PDs and EPDs.    
    We compute persistent homology (PH) and extended persistent homology (EPH) via the sequences \eqref{SI:eq:persmod} and \eqref{SI:eq:extpersmod} using the horizontal and vertical filtration 
    % gives two EPDs containing persistence pairs $(b, d)$ which correspond to the birth and death of topological features within the concatenated sequence.
    % The values $b$ and $d$ correspond to the location (in co-ordinate directions) of each feature and the persistence $d-b$ gives the size.
    illustrating the additional information offered by EPH.
    As the final step in computing topological summary statistics, we vectorise EPDs into Persistence Images and persistence statistics according to section \ref{SI:subsec:vectorisations}.
    The left panel of A was created in \url{https://BioRender.com}
    }
     \label{SI:fig:topfeatures}
\end{figure}\clearpage

\begin{figure}
    \centering
    \includegraphics[width=.8\linewidth]{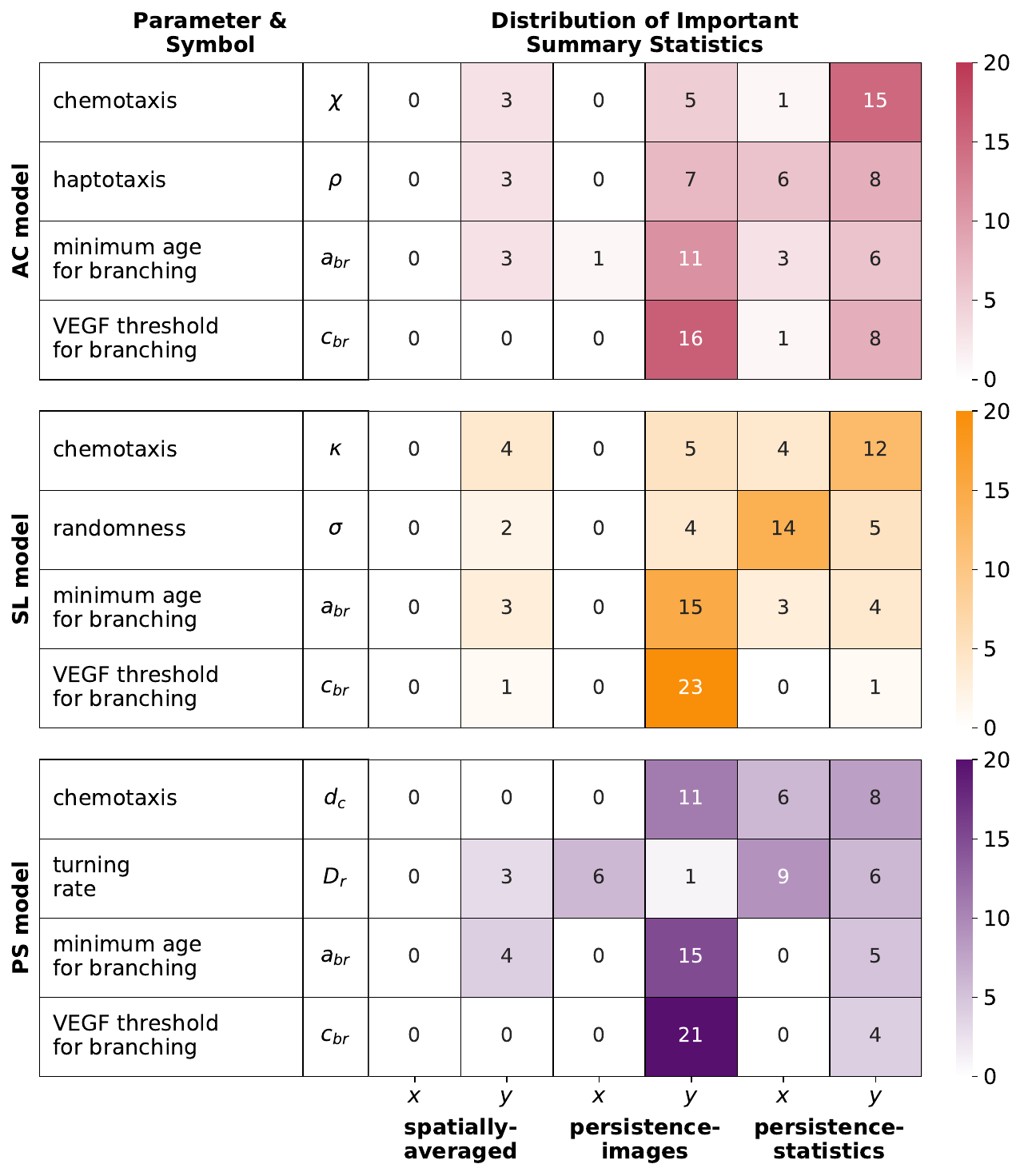}
    \caption{\textbf{Distribution of important summary statistics.} 
    Defining importance as in section 3.1 as the mean decrease in L2 impurity \eqref{SI:eq:RFl2} induced by each summary statistic in each parameter's Random Forest, we report the types of the top $100$ summary statistics.
    In Figure 1 of the main text, we disaggregate the important summary statistics by type (spatially-averaged or topological) and direction (horizontal (x) and vertical (y)). 
    Here, we further delineate and show how many of each summary statistic comes from each topological vectorisation--Persistence Images and persistence statistics.
    No spatially-averaged summary statistics computed in the $x$ direction appear among the top $100$ for any parameter.
    \cite{vectorisationSurvey} found that persistence statistics perform better than other vectorisation methods (including Persistence Images) on common classification tasks. 
    We find that a mixture of Persistence Images and persistence statistics are useful for the inference of parameters in each model.}
    \label{SI:fig:top_ss}
\end{figure}
\clearpage
\section{ABC and Random Forest Methodology}\label{SI:sec:RF}
We use Random Forests \citep{orginalRF} to choose informative summary statistics, which we then use to infer spatial parameters and select between spatial models using ABC.
In this section, we give further details on each step described in section 3 of the main text.
\subsection{Step 1: Identify Informative Summary Statistics For Each Model}\label{SI:subsec:step1}
ABC requires a distance function $\nu: \mathcal{D} \times \mathcal{D} \to \mathbb{R}$ to measure the discrepancy between observed data and simulated data. In
section \ref{SI:sec:ss} we showed how to compute spatially-averaged and topological summary statistics from angiogenesis datasets.
Combining all spatially-averaged and topological summary statistics into a single vector gives a total of $n_\text{f}=1108$ features which may be computed from an angiogenesis dataset.
We now identify an informative subset of these summary statistics which we will use to construct the distance function $\nu$.

We train one regression Random Forest per parameter per angiogenesis model to learn the relationship between parameter values and summary statistics.  
By training one Random Forest per parameter, we aim to identify the summary statistics that capture the effect of varying each individual parameter's value on simulated data.
To generate training data, we draw $n=10,000$ parameter values $\theta_i$ uniformly from the parameter ranges given in Table \ref{SI:tab:models_params}.
We generate an angiogenesis dataset corresponding to each parameter value by simulating each model up to time $t_\text{final}=4$ and consider the angiogenesis dataset at its final time-step as described in section \ref{SI:sec:models}.
We compute spatially-averaged and topological summary statistics to form a vector $X_i$ of summary statistics corresponding to the parameter value $\theta_i$, which together form training data $\mathcal{X}$ and $\mathcal{Y}$.

To construct decision trees that make up each Random Forest, we take bootstraps of $n_\text{samples}=2,000$ pairs $(X_i, \theta_i)$ sampled independently (with replacement) from the training data.
Each decision tree is made up of internal nodes which repeatedly partition the training data and leaf nodes which contain $n_\text{min}=5$ or fewer pairs from the bootstrap.
At an internal node $\mathcal{N}$, $n_\text{f} / 3$ features are randomly selected and considered to create a splitting rule.
Splitting rules are conditions of the form $X_i^j > r$ for some co-variate $j$ and splitting bound $r$.
Pairs $(X_i, \theta_i) \in \mathcal{N}$ for which a condition is satisfied are allocated to the right daughter node $\mathcal{N}_\text{R}$ and the others are passed to left daughter node $\mathcal{N}_\text{L}$.
The bootstrap of the training data is repeatedly partitioned by internal nodes in this way until $n_\text{min}=5$ or fewer pairs $(X_i, \theta_i)$ are allocated to a node, wherein it becomes a leaf node.

To decide the co-variate index $j$ and splitting bound $r$ used at each internal node, Random Forests consider a loss function with the general form of Equation \eqref{SI:eq:RFloss}.
Random Forests choose co-variates and splitting bounds which minimise $\Delta_\text{loss}$ within each internal node.
\begin{equation}
    \Delta_\text{loss} = \frac{\abs{\mathcal{N}_L}}{\abs{\mathcal{N}}}  
    Q(\mathcal{N}_L) 
    + 
    \frac{\abs{\mathcal{N}_R}}{\abs{\mathcal{N}}}
    Q(\mathcal{N}_R) 
\label{SI:eq:RFloss}
\end{equation}

$Q(\mathcal{N})$ is a measure of the impurity of samples allocated to node $\mathcal{N}$ and $\abs{\mathcal{N}}$ is the number of samples allocated to the node. 
As is common in regression Random Forests, we use the L2 impurity given by Equation \eqref{SI:eq:RFl2} to decide splitting rules. 
The L2 purity measures the variance in parameter values from the mean $\bar{\theta}_\mathcal{N}$ among pairs allocated to the same daughter node.
\begin{equation}
    Q(\mathcal{N}) =
    \sum_{\theta_i: (X_i, \theta_i) \in \mathcal{N}} (\theta_i - \bar{\theta}_{\mathcal{N}})^2
\label{SI:eq:RFl2}
\end{equation}

To choose the number $n_\text{tree}$ of decision trees to use in each Random Forest, we follow the advice in \cite{ABCRF_paraminf} and compute the out-of-bag mean square error of Random Forests constructed with different numbers of trees.
We note in Figure \ref{SI:fig:ntree1} that the out-of-bag error decreases when adding additional trees, but that this improvement is small when using more than $100$ trees in the Random Forest for each parameter in each model.
We therefore use $n_\text{tree}=100$.

To select a subset of the spatially-averaged and topological summary statistics for use in subsequent ABC algorithms, we consider the importance of each feature in the Random Forest for each parameter.
The importance of a co-variate $j$ in a trained Random Forest is defined as the mean decrease in impurity $Q(\mathcal{N}) - (Q(\mathcal{N}_L) + Q(\mathcal{N}_R))$ achieved by all internal nodes $\mathcal{N}$ which use $j$ in their splitting rule.
Features with high importance are therefore those which are most effective in partitioning the training data, meaning they capture the effect of model parameters on model simulations.

Ranking the $n_\text{f}=1108$ summary statistics by importance, we observe in Figure \ref{SI:fig:justify_100f} that a small number of features carry most of the predictive power for each parameter in each model.
To construct the distance function $\nu$ used in ABC, we wish to use those statistics which distinguish data simulated using different model parameters and omit those which do not.
Therefore, we cycle through each parameter in each model and choose the top $25$ most important summary statistics for each parameter that have not already been selected, giving a list of $100$ features in total for each model.
In general, we recommend computing the importance of a long-list of summary statistics and choosing $n_\text{s}$ so that statistics with low importance are omitted. 
\newpage
\begin{figure}[!ht]
    \centering
\includegraphics[width=0.75\linewidth]{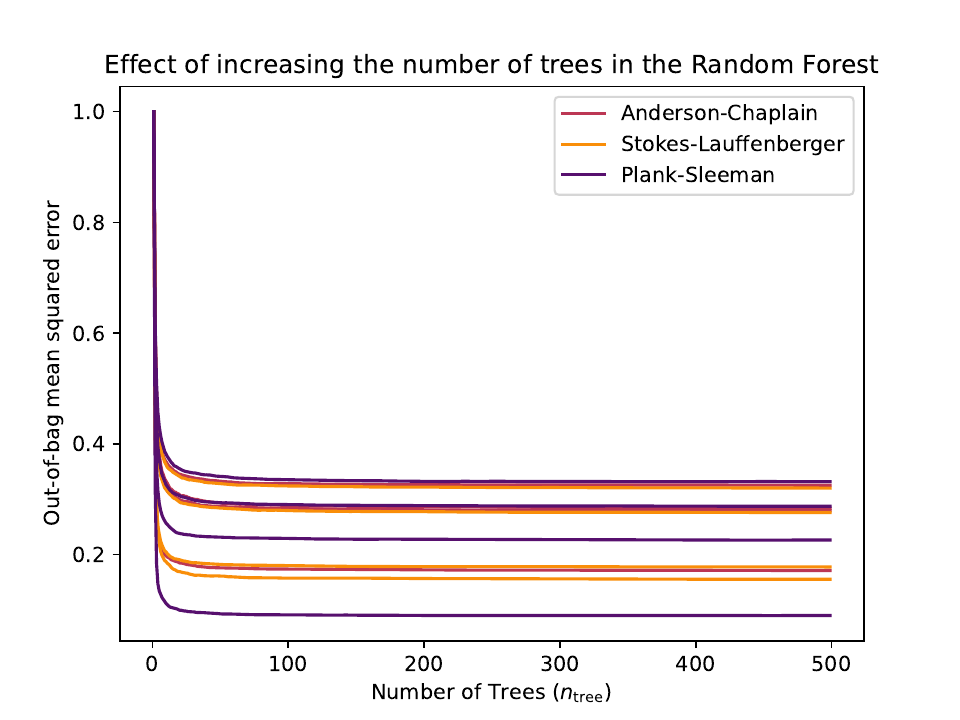}
    \caption{The out-of-bag mean square error of regression Random Forests learning the relationship between summary statistics $X_i$ and parameter values $\theta_i$ (scaled by its initial value).
    The out-of-bag mean square error initially decreases, but negligibly after $n_\text{tree} = 100$.}
    \label{SI:fig:ntree1}
\end{figure}
\newpage
\subsection{Step 2: Fit Each Model To The Observed Data}
For observed data $\mathcal{D}^*$, we approximate the parameter posterior $p(\Theta|\mathcal{D}^*)$ using Approximate Bayesian Computation (ABC).
ABC algorithms follow the basic format of Algorithm \ref{ABC-rej}, which is ABC with rejection sampling.
In this section, we discuss the choices of priors $p(\Theta)$, distance function $\nu$, and tolerance $\epsilon$.
\begin{figure}[h]
\centering
\begin{minipage}{0.475\linewidth} 
\begin{algorithm}[H]
    \caption{ABC with rejection sampling}
  \begin{algorithmic}[1]
  
    \INPUT Observed data $\mathcal{D}^*$, a model which generates data $\mathcal{D}$ from parameters $\Theta$ with prior $p(\Theta)$, a distance function $\nu: \mathcal{D} \times \mathcal{D} \to \mathbb{R}$ comparing simulated and observed data, and a small tolerance $\epsilon >0$.
    \OUTPUT A collection of samples $\theta_i$ from the posterior $p(\Theta | \mathcal{D}^*)$
    \FOR{Candidate parameter values $\theta_i$ sampled from $p(\Theta)$}
    \STATE{Simulate data $\mathcal{D}_i$ from the model using parameter value $\theta_i$}
     \IF{$\nu (\mathcal{D}^*, \mathcal{D}_i) \leq \epsilon$}
    \STATE{accept $\theta_i$}
\ENDIF
\ENDFOR
  \end{algorithmic}
      \label{ABC-rej}
\end{algorithm}
\end{minipage}
\end{figure}

\noindent\textbf{Choosing priors $p(\Theta)$}\\
For each parameter in each model, we use the uniform prior $p(\theta) = \mathcal{U}_{[\theta_\text{min}, \theta_\text{max}]}(\theta)$, where $[\theta_\text{min}, \theta_\text{max}]$ are the parameter ranges in Table \ref{SI:tab:models_params}.
Each parameter range is either taken from existing literature or chosen with the aim of exhibiting a wide range of simulation behaviour in each model--see Table \ref{SI:tab:models_params} for details.\\

\noindent\textbf{Choosing distance function $\nu$}\\
Following section \ref{SI:subsec:step1}, we choose a subset of $n_\textnormal{s} = 100$ summary statistics for each model from among the spatially-averaged and topological summary statistics.
Due to the computational demands of parameter inference using ABC-SMC, it is not computationally feasible to benchmark multiple potential values of $n_\textnormal{s}$ or multiple subsets of summary statistics.
We therefore advocate a heuristic approach to choosing he number of summary statistics to use in $\nu$.
As shown in Figure \ref{SI:fig:justify_100f}, most of the predictive power of each Random Forest lies in the top $100$ summary statistics.
Including additional summary statistics is unlikely to aid parameter inference, as they would only add noise to the ABC distance function $\nu$.
Since it is likely that different summary statistics will be informative for different models, the value of $n_s$ must be set large enough to ensure that some of the same summary statistics are chosen for all three models.
Setting $n_\textnormal{s}=100$ results in $\tilde{n}_\text{s}=30$ summary statistics selected for all three models, which we deem sufficient to train the Random Forests needed for model selection.

It is possible that the important summary statistics may be poorly scaled.
For example, connected components are generally larger than loops in the angiogenesis datasets we consider.
Small differences in the size and location of loops in angiogenesis datasets may be stronger predictors of parameter values than the size and locations of connected components, yet this may not be reflected in the distance function $\nu$ if the corresponding topological summary statistics are left unscaled.
We therefore construct a scaling function which divides each summary statistic by the maximum absolute value of that summary statistic among the training data for each model.
The ABC distance function is then $\nu(\mathcal{D}^*, \mathcal{D}_i) = \norm{x^* - x_i}_2$ where $\mathcal{D}^*$ is observed data, $\mathcal{D}_i$ is simulated data, and $x^*$ and $x_i$ are the $100$ summary statistics with the scaling applied.
When $\mathcal{D}^*$ comprises several instances of observed data, we compute the distance $\nu$ for each observed dataset and take their average.\\

\noindent\textbf{Choosing tolerance $\epsilon$}\\
Using Algorithm \ref{ABC-rej} with a single value of the tolerance $\epsilon$ may lead to a poor approximate posterior or slow convergence.
If $\epsilon$ is too large, too many parameters will be accepted from the prior, and if $\epsilon$ is too small, few parameters will be accepted and many simulations will be needed to approach the true posterior.
Instead, we use the ABC-SMC algorithm of \cite{smcDelMoral} which outputs a series of $n_\text{pop}$ intermediate distributions, which correspond to a decreasing sequence of tolerances $\epsilon_0> \dots> \epsilon_{n_\text{pop}}>0$.
Beginning with a population of $N_\text{pop}$ parameters sampled from a prior distribution, the initial tolerance $\epsilon_0$ is chosen so that a predetermined fraction $\alpha \in (0, 1)$ of parameters are accepted.
The effective sample size (ESS) of each population is a measure of the independence of parameters within a population.
\cite{smcDelMoral} computes the ESS of each subsequent population and chooses further tolerances such the ESS of each population decreases by $\alpha$.
We use $\alpha=0.8$ and generate $n_\text{pop}=50$ populations, each containing $N_\text{pop}=1,000$ parameters, and take the final population as the approximate ABC-SMC posterior $p(\Theta|\mathcal{D}^*)$.

\newpage

\subsection{Step 3: Approximate The Model Posterior}
We approximate the model posterior $p(m | \mathcal{D}^*)$ for observed data $\mathcal{D}^*$ by training two more Random Forests, following the method of \cite{ABCRF_modelchoice}.
Using the training data $(X_i, m_i)$ obtained from the training data used in step 1 by replacing the parameter $\theta_i$ with the model index $m_i \in \{\text{AC}, \text{SL}, \text{PS}\}$, we train a classification Random Forest to learn the relationship between summary statistics $X_i$ and model index $m_i$.

Step 1 chooses a subset of the spatially-averaged and topological summary statistics which are important for each for each model.
Since the importance of each summary statistic to a model is determined by that model's training data, it is possible that different summary statistics are important for different models.
Using summary statistics which are only informative for a single model to approximate the model posterior may bias predictions in favour of that model.
We therefore learn the relationship between summary statistics and model index using only those summary statistics which are important for all angiogenesis models.

Specifically, we modify $X_i$ to contain only those $\tilde{n}_\text{s} \leq 100$ features which appear in the top $100$ summary statistics for all three models.
We found that $\tilde{n}_\text{s} = 30$ summary statistics appear in the top $100$ for all angiogenesis models, and hence the $X_i$ are modified to include only these entries.
Figure \ref{SI:fig:justify_100f} shows how the number of summary statistics that appear in the top $n_\text{s}$ summary statistics for all models grows as $n_\text{s}$ increases.
We also plot the expected number of common summary statistics among the three models if summary statistics were selected randomly.
$n_\text{s}$ should be large enough to include some common summary statistics that important for all models, but not so large as to allow the possibility that some summary statistics are chosen for all three models by random chance.
The expected number of common summary statistics if $n_\text{s}=100$ are chosen randomly for each model is approximately $1$, which suggests that the $\tilde{n}_\text{s} = 30$ common summary statistics are indeed informative for all three models, and were not chosen by random chance.

As in step 1, we bootstrap $n_\text{samples}=2,000$ pairs $(X_i, m_i)$ to train $n_\text{tree}=100$ decision trees, randomly choosing $\tilde{n}_\text{s}/3$ summary statistics at each internal node to consider for splitting rules, and terminating at leaf nodes only when $n_\text{min}=5$ samples remain.
To decide on the co-variate index $j$ and splitting bound $r$ in the partition rule $X_i^k < r$ at each internal node, we use the Gini impurity defined in Equation \eqref{SI:eq:RFgini}.
For a node $\mathcal{N}$, let $p_i$ be the proportion of pairs $(X_i, m_i)$ allocated to $\mathcal{N}$.
 The Gini impurity is a measure of how many different model indices are allocated to node $\mathcal{N}$, which the splitting rule minimises.
\begin{equation}
    Q(\mathcal{N}) =
    \sum_{m_i \in \{\text{AC}, \text{SL}, \text{PS}\}}p_i(1-p_i)  
\label{SI:eq:RFgini}
\end{equation}
The trained classification Random Forest provides a prediction $RF(X^*)$ of the model $m^*$ which generated observed data $\mathcal{D}^*$ (which we aggregate when $\mathcal{D}^*$ contains multiple instances of observed data).
The Random Forest also provides an out-of-bag error for each pair $(X_i, m_i)$, which is the proportion of those decision trees which did not use this pair in their training bootstrap which predict the incorrect model index $m_i$.
Following \cite{ABCRF_modelchoice}, we train another regression Random Forest using the same training data to learn the relationship between summary statistics $X_i$ and this mis-classification error rate $p(RF(X_i) \neq m_i)$ of the first Random Forest.
Predicting the out-of-bag error $p(RF(X^*) \neq m^*)$ for the observed data $\mathcal{D}^*$ gives an estimate $1-p(RF(X^*) \neq m^*)$ for the model posterior $p(m=m^* | \mathcal{D}^*)$.\\
\newpage

\begin{figure}[H]
    \centering
    \includegraphics[width=0.75\linewidth]{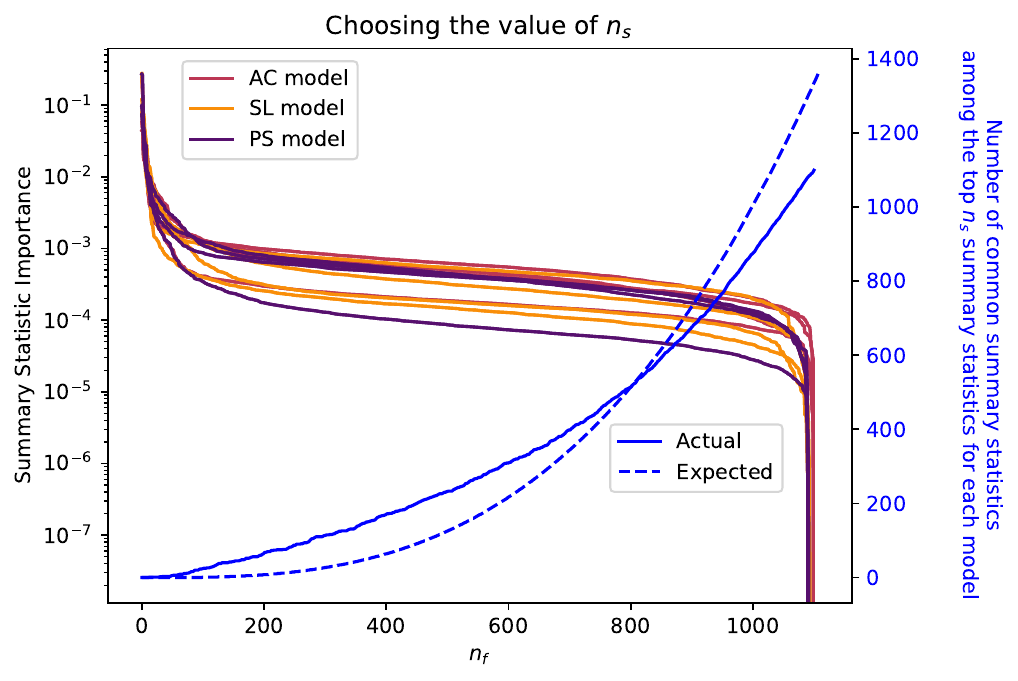}
    \caption{Left axis: for each parameter in each model, we rank summary statistics by their importance and observe how importance decreases. Right axis: we compare how many summary statistics appear among the top $n_\text{s}$ most important features for all models as $n_\text{s}$ is increased and plot the expected number of common summary statistics if they were chosen randomly.
    Choosing $n_\text{s} = 100$ includes approximately the $25$ most important summary statistics for each parameter.
    Selecting more summary statistics than this would include those which are of little use in capturing the effect of parameter values on simulated data.
    Choosing $n_\text{s}=100$ also ensures that $\tilde{n}_\text{s}=30$ summary statistics appear in the top $n_\text{s}$ most important summary statistics for all models.
    If $100$ summary statistics were chosen uniformly randomly for each model, the expected number that would be common to all three models is approximately $1$.
    }
    \label{SI:fig:justify_100f}
\end{figure}
\newpage
\section{Toy Model Example}\label{SI:sec:toymodelexample}
\subsection{Two Two-Parameter Toy Models}
We verify our pipeline for parameter inference and model selection on two toy models which can simulate angiogenesis datasets similar to the simple vascular network in Figure \ref{SI:fig:topfeatures}.
While the likelihood functions for the angiogenesis models analysed in the main text are unavailable, we construct simple toy models from which we can derive a likelihood function.
Using appropriate priors and Bayes' rule, we can then derive the exact parameter posterior for each toy model, and compare it to the approximate parameter posterior produced using our pipeline (steps 1-2 of section 3 of the main text).
We then test our method of model selection (step 3 of section 3 of the main text) on the two toy models.

Both toy models initialise two tip EC along the bottom of a the square domain $\mathcal{I} = [0, 1]^2$, which move diagonally upwards.
The tip ECs anastomose, branch, and then anastomose again, forming a central loop before reaching the VEGF source at the top of the domain.
We derive two stochastic spatial models from this simple construction by parametrising the size and shape of the central loop. 
The first model, Toy-Circle (TC), has two parameters $r$ and $c$ which define the centre and radius of a circular loop in the middle of the domain.
Given a parameter pair $(r, c)$, the TC model produces a blood vessel network whose central loop is a circle of radius $r+\varepsilon$ and centre $(0.5, c+\varepsilon)$, where each $\varepsilon$ is identically and independently drawn from $\mathcal{U}_{[-0.03, 0.03]}$, the uniform distribution between $-0.03$ and $0.03$.
The randomly sampled $\varepsilon$ are added to each parameter to introduce random variation 
into the model which is simple enough to allow derivation of the model's likelihood function.
The second model, Toy Ellipse (TE), uses a similar construction to produce a vascular network.
However, the TE model uses parameters $a$ and $b$, to define the size of the horizontal radius and vertical radius of an ellipse which makes up the loop in the centre of the domain.
Given a parameter pair $(a, b)$, the TE model produces a blood vessel network whose central loop is an ellipse centred at $(0.5, 0.5)$ with horizontal radius $a+\varepsilon$ and vertical radius $b+\varepsilon$.
The noise $\varepsilon$ is sampled in the same way as in the TC model.
Each toy model, including the range of each parameter, is summarised in Table \ref{SI:tab:toymodels}.

\subsection{Computing The Exact Likelihood and Posterior}
Suppose an angiogenesis dataset $\mathcal{D}_i$ is simulated from the TC model using parameters $(r, c)$ and has radius $r_i$ and centre $c_i$. 
The likelihood of $\mathcal{D}_i$ is the product of uniform distributions: $p(\mathcal{D}_i|r, c) = p(r_i, c_i | r, c) = \mathcal{U}_{[r-0.03, r+0.03]}(r_i) \times \mathcal{U}_{[c - 0.03, c+0.03]}(c_i)$.
Assume that uniform priors $p(r) = \mathcal{U}_{[r_\text{min}, r_\text{max}]}(r)$ and $p(c) = \mathcal{U}_{[c_\text{min}, c_\text{max}]}(c)$ are used for $r$ and $c$ respectively, with the maximum and minimum value of each parameter taken from Table \ref{SI:tab:toymodels}.
Given observed data $\mathcal{D}^* = \{ \mathcal{D}^*_1, \dots, \mathcal{D}^*_{n^*} \}$ consisting of $n^*$ angiogenesis datasets whose central loops have centres $c_1^*, \dots, c_{n^*}^*$ and radii $r_1^*, \dots, r_{n^*}^*$, the parameter posterior can be computed exactly by  \eqref{SI:eq:TCpostbayes}--\eqref{SI:eq:TCpost}.
\begin{align}
p(r, c | \mathcal{D}^*) =
\prod_{i=1}^{n^*} p(r, c | D^*_i) 
&\propto \prod_{i=1}^{n^*} p(\mathcal{D}_i^* | r, c) \times p(r, c) \label{SI:eq:TCpostbayes} \\
&= \prod_{i=1}^{n^*} p(r^*_i, c^*_i| r, c) \times p(r) \times p(c)\\
&= \prod_{i=1}^{n^*} \big[
\mathcal{U}_{[r-0.03, r+0.03]}(r_i^*) \times 
\mathcal{U}_{[c-0.03, c+0.03]}(c_i^*)  \big]
\times \mathcal{U}_{[r_\text{min}, r_\text{max}]}(r)
\times \mathcal{U}_{[c_\text{min}, c_\text{max}]}(c) \\
&= \prod_{i=1}^{n^*} \big[
\mathcal{U}_{[r^*_i-0.03, r^*_i+0.03]}(r) \times 
\mathcal{U}_{[c^*_i-0.03, c^*_i+0.03]}(c)  \big]
\times \mathcal{U}_{[r_\text{min}, r_\text{max}]}(r)
\times \mathcal{U}_{[c_\text{min}, c_\text{max}]}(c) \\
\begin{split}
&=\mathcal{U}_{[\max\{r_\text{min}, \max_{i=1}^{n^*} r^*_i - 0.03\}, \min\{r_\text{max}, \min_{i=1}^{n^*} r^*_i + 0.03]\}}(r) \times \\
    &\qquad\mathcal{U}_{[\max\{c_\text{min}, \max_{i=1}^{n^*} c^*_i - 0.03\},
    \min\{c_\text{max}, \min_{i=1}^{n^*} c^*_i + 0.03])\}}(c)
    \label{SI:eq:TCpost}
\end{split}
\end{align}
Line \eqref{SI:eq:TCpostbayes} uses Bayes' rule and factors out the evidence $p(\mathcal{D}_i^*)$.
We assume throughout that $r_i^*$ and $c_i^*$ have values within the ranges given in Table \ref{SI:tab:toymodels} (plus or minus 0.03) and are within $0.06$ of each other (otherwise the likelihood and posterior are both 0).
For the TE model, assuming that observed data $\mathcal{D}^* = \{\mathcal{D}_1^*, \dots, \mathcal{D}_{n^*}^*\}$ is a collection of angiogenesis datasets where the central loops are ellipses with horizontal radii $a_1^*, \dots, a_{n^*}^*$ and vertial radii $b_1^*, \dots, b_{n^*}^*$, a similar computation gives the parameter posterior $p(a, b| \mathcal{D}^*)$ as \eqref{SI:eq:TEpost}. 
\begin{equation}
\begin{split}
p(a, b | \mathcal{D}^*) &= 
    \mathcal{U}_{[\max\{a_\text{min}, \max_{i=1}^{n^*} a^*_i - 0.03\}, \max\{a_\text{max}, \min_{i=1}^{n^*} a^*_i + 0.03\}]}(a) \times \\
    &\qquad\mathcal{U}_{[\max\{b_\text{min}, \max_{i=1}^{n^*} b^*_i - 0.03\},
    \min\{b_\text{max}, \min_{i=1}^{n^*} b^*_i + 0.03]\}}(b)
    \label{SI:eq:TEpost}
\end{split}
\end{equation}
\begin{table}[]
\centering
\begin{tabular}{|ll||ll|}
\hline
\multicolumn{2}{|c||}{\textbf{Toy-Circle (TC) model}} & \multicolumn{2}{c|}{\textbf{Toy-Ellipse (TE) model}} \\ \hline
\multicolumn{2}{|c||}{parameters} & \multicolumn{2}{c|}{parameters} \\ \hline
\multicolumn{1}{|l|}{\begin{tabular}[c]{@{}l@{}}Name and\\ Symbol\end{tabular}} & Range & \multicolumn{1}{l|}{\begin{tabular}[c]{@{}l@{}}Name and\\ Symbol\end{tabular}} & Range \\ \hline
\multicolumn{1}{|l|}{\begin{tabular}[c]{@{}l@{}}Loop radius\\ $r$\end{tabular}} & $[0.07, 0.23]$ & \multicolumn{1}{l|}{\begin{tabular}[c]{@{}l@{}}Loop horizontal radius\\ $a$\end{tabular}} & $[0.07, 0.43]$ \\ \hline
\multicolumn{1}{|l|}{\begin{tabular}[c]{@{}l@{}}Loop center\\ $c$\end{tabular}} & $[0.34, 0.66]$ & \multicolumn{1}{l|}{\begin{tabular}[c]{@{}l@{}}Loop vertical radius\\ $b$\end{tabular}} & $[0.07, 0.43]$ \\ \hline
\end{tabular}\\

\caption{The parameters of each toy model. 
Given parameter pair $(r, c)$, the TC model simulates a network whose central loop is a circle with radius $r+\varepsilon$ and center $c+\varepsilon$, where each $\varepsilon$ is sampled from $\mathcal{U}_{[-0.03, 0.03]}(\varepsilon)$ independently and identically. 
The TE model uses the parameter pair $(a, b)$ to simulate a network whose central loop is an ellipse with horizontal radius $a+\varepsilon$ and vertical radius $b+\varepsilon$ with each $\varepsilon$ also sampled from $ \mathcal{U}_{[-0.03, 0.03)]}(\varepsilon)$.}
\label{SI:tab:toymodels}
\end{table}
\subsection{Parameter Inference and Model Selection}
We generate four synthetic test-cases for each toy model by choosing parameter pairs which cover a range of circles/ellipses.
We generate ${n^*}=2$ simulations at each test parameter pair and use them as observed data $\mathcal{D}^*$.
Following steps 1-2 of section 3 of the main text, we approximate the parameter posteriors $p(r, c | \mathcal{D}^*)$ for the TC model and $p(a, b|\mathcal{D}^*)$ for the TE model in each test-case.

In Figure \ref{SI:fig:toyparaminf}, we plot the true posterior, calculated using Equations \eqref{SI:eq:TCpost} and \eqref{SI:eq:TEpost}, as a light blue square, and the approximate posterior resulting from our pipeline in dark blue.
The true parameter value falls within the true posterior, but not necessarily at its centre, since the true posterior depends on ${n^*}=2$ simulations of the toy model (used as observed data) which contain random noise.
In all four test-cases for each model, the approximate posterior closely matches the true posterior.

To test our method of spatial model selection, we perform step 3 of section 3 of the main text and approximate the model posterior $p(m|\mathcal{D}^*)$ for the same four test-cases, giving results in Figure \ref{SI:fig:toymodelselection}.
Since test-cases 1 and 2 of the TC model contain circles not centred at $(0.5, 0.5)$, and test-cases 1 and 2 of the TE model contain ellipses with unequal horizontal and vertical radii, only the true model can produce their observed data.
In these test-cases, we successfully infer the true model and correctly approximate posterior probabilities as $1$ for the true model and $0$ for the other model.
The observed data in test-cases 3 and 4 of each toy model, however, could have been generated by either model.
In these test-cases, the central loop in the observed data is is either a circle with centre $c=(0.5, 0.5)$, or an ellipse with equal horizontal and vertical radii, which either toy model can reproduce.
We therefore expect a non-zero posterior probability for each model in these test-cases, and we successfully approximate this.
The approximate model posterior is still able to identify the true model in these cases, and we estimate only a small posterior probability of the incorrect mode in each test-case.
While both models can simulate data similar to the observed data in each test-case, the true model will do so more often (for more parameter values).
The training data from which we learn the relationship between summary statistics and model index therefore contains more simulations similar to the observed data when the true model is used, which may be why it predicts the true model with higher probability.
\newpage

\begin{figure}[!ht]
    \centering
        \hspace{1cm}
    \includegraphics[width=.275\linewidth]{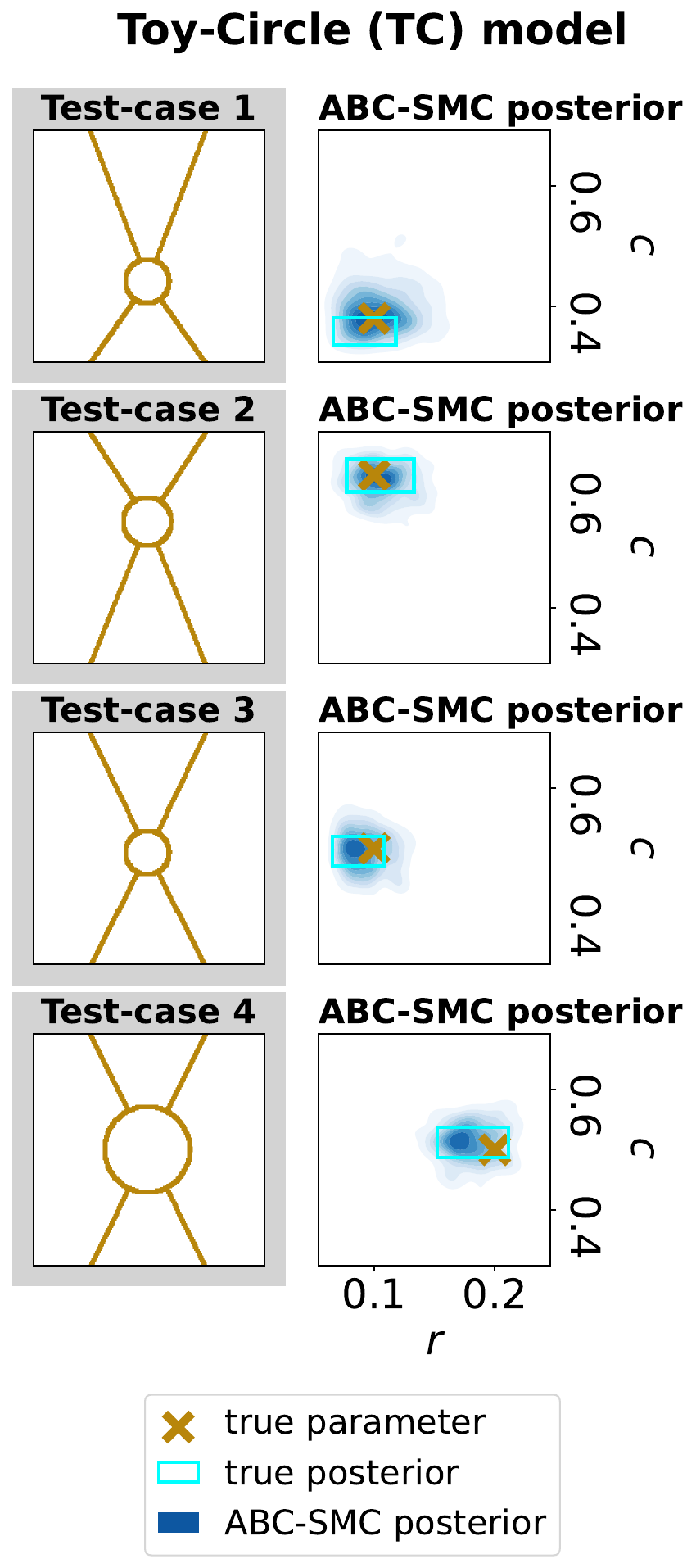}
    \hfill
     \includegraphics[width=.275\linewidth]
     {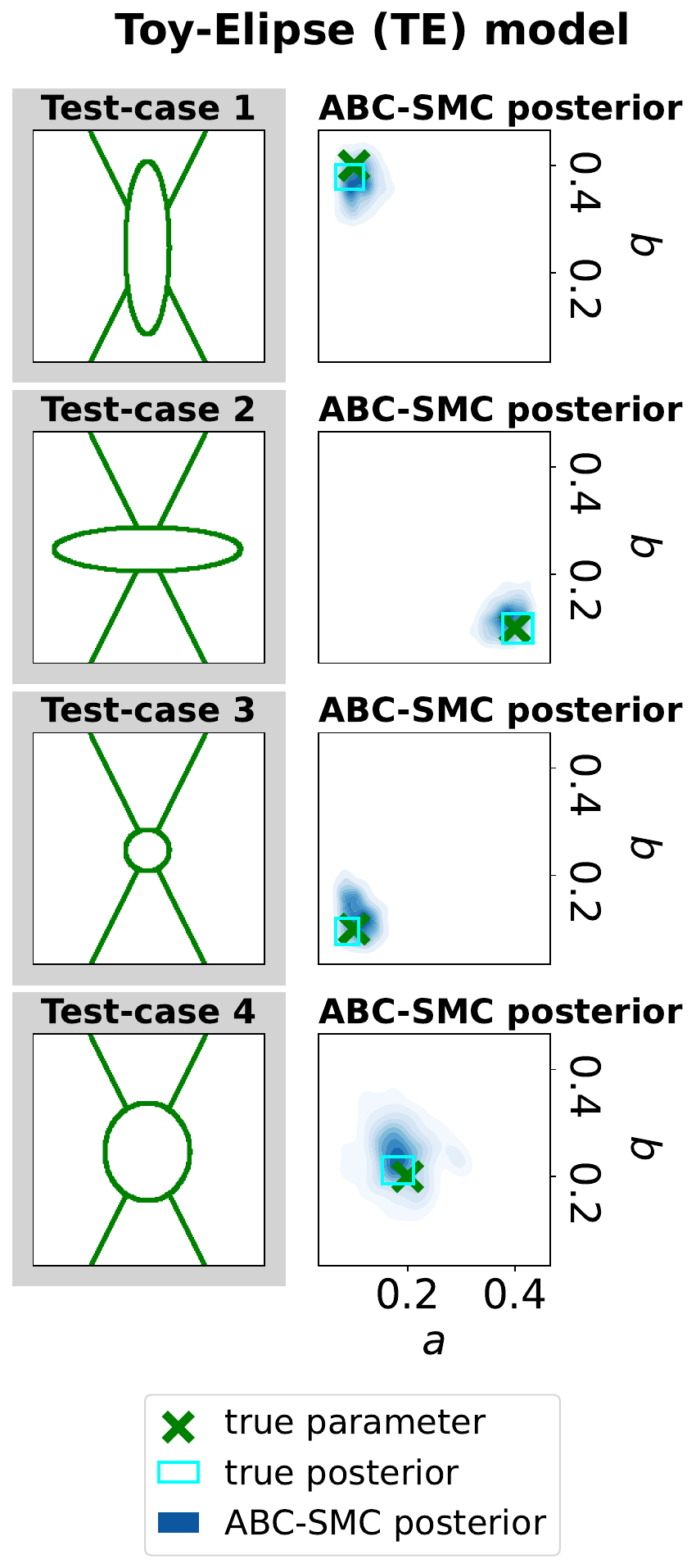}
        \hspace{1cm}
    \caption{
    Parameter inference in the TC and TE models.
    To generate observed data, we fix a set of parameter values and simulate the model in question ${n^*}=2$ times.
For each model, we pick four parameter sets to generate four test-cases and show one example of the angiogenesis datasets they simulate.
We then use steps 1-2 of section 3 to find informative summary statistics which we use to fit each model to the observed data in each test-case. 
The output of ABC-SMC is a population of parameter values which approximate the parameter posterior $p(\Theta | \mathcal{D}^*)$.
We plot the resulting distributions (fitting a Gaussian kernel to the parameter values accepted in the final population of the ABC-SMC algorithm).
We also plot the true parameter which generated the observed data and the true parameter posterior, which we can compute exactly since the toy models are simple.}
    \label{SI:fig:toyparaminf}
\end{figure}
\newpage

\begin{figure}[!ht]
    \centering
    \includegraphics[width=0.45\linewidth]{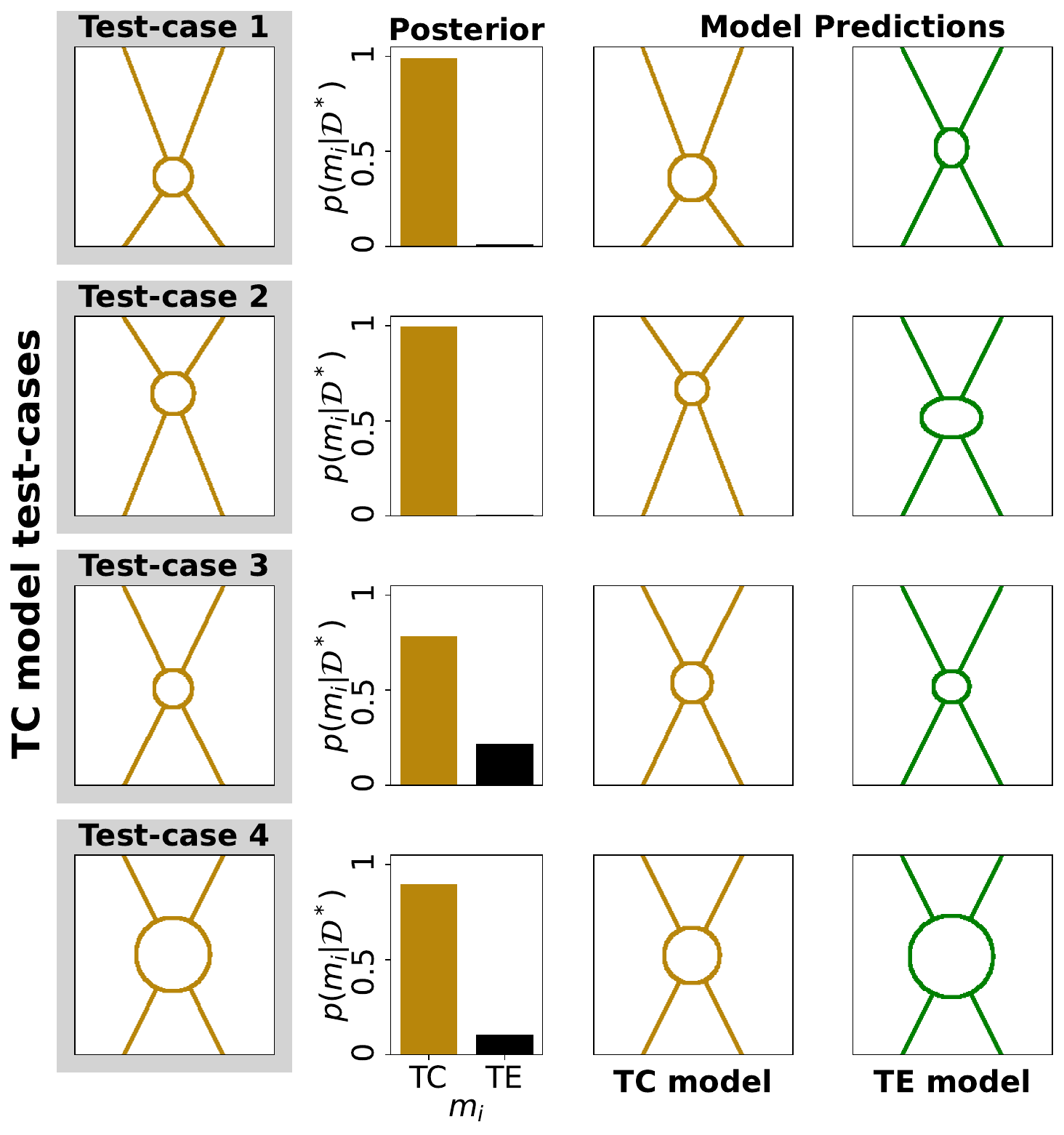}
    \hfill
     \includegraphics[width=0.45\linewidth]{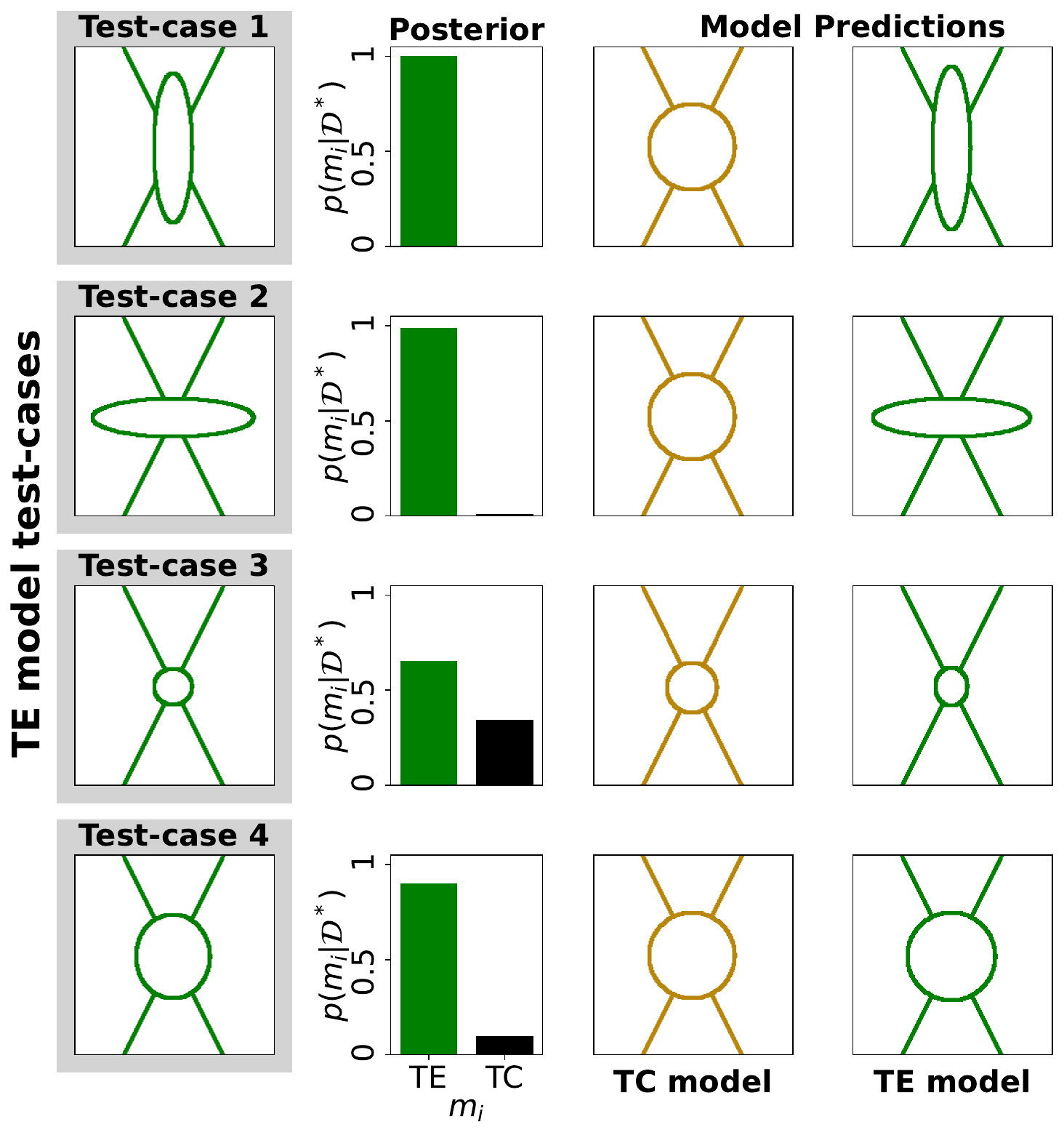}

    \caption{Model selection among the TC and TE toy models.
    We use the same test-cases as in Figure \ref{SI:fig:toyparaminf} in which we generated observed data $\mathcal{D}^*$ by simulating each model at known parameter values.
    Again we show one example of the angiogenesis datasets they simulate.
    Performing step 3 of section 3 of the main text gives an approximation of the model posterior $p(m|\mathcal{D}^*)$, which we show for each test-case.
    The approximate posterior selects the correct model in each test-case, giving the other model 0 posterior probability in test-cases 1 and 2 (where the observed data could only have been generated by the true model) and non-zero posterior probability in test-cases 3 and 4 (where either model could have generated the observed data).
    We fit each model to the observed data using step 2 of section 3 of the main text, and show one simulation of each model using a parameter value sampled from the approximate parameter posterior for each model.
    This prediction represents each model's best approximation to the observed data.
    In test-cases 1 and 2, only the true model predicts data similar to the observed data, whereas in test-cases 3 and 4, both models produce a visually similar approximation.}
    \label{SI:fig:toymodelselection}
\end{figure}

\newpage

\section{Supplementary Results}\label{SI:sec:suppresults}

We experiment with different values of $t_\text{final}$ to investigate whether our pipeline can infer parameters and select the correct model with less simulation information.
Specifically, we modify the AC, SL, and PS models so that they terminate after times $t_\text{final}=2.5, 2.5, 1.5$ respectively, meaning that ECs only move around half of the way to the tumour within the simulation.

Figure \ref{SI:fig:suppparaminf} shows the results of parameter inference using the same test-cases as Figure 3 of the main text (but with simulation times decreased).
Interestingly, the non-branching parameters in each model are inferred as well as, if not better than, when the simulations are run for the full time period.
The approximate posterior for these parameters are narrower than in Figure 3 of the main text.
When the simulation time is shorter, the movement parameters (chemotaxis in all models, and either haptotaxis, randomness and turning rate) appear to have a stronger effect on the simulated data.
For example, the value of the chemotaxis parameter determines how quickly ECs travel towards the tumour.
In simulations which run up to $t_\text{final}=4$, most ECs reach the tumour.
However when time is limited, the chemotaxis parameter affects how close ECs get to the tumour, which may be why its value is inferred with more certainty when simulation time is limited.

The branching parameters $a_\text{br}$ and $c_\text{br}$ are inferred less accurately (and approximate parameter posterior distributions are wider) when simulations are time-limited, and are in some cases not inferred correctly.
Perhaps unsurprisingly, when simulations are run for shorter times, the ECs exhibit less branching behaviour, which may explain why these parameters are now more difficult to infer.

Figure \ref{SI:fig:suppmodelselection} attempts model selection, again using the same (modified) test-cases as Figure 4 of the main text.
In each test-case, the correct model is identified, despite the simulations being time-limited.
However, the posterior probability of the incorrect models is higher, and it appears to be more difficult to identify the correct model when less simulation information is available.
The `model predictions' appear to suffer from the shorter simulation time.
Since some parameters (in particular, those which regulate branching) are now uncertain, parameters sampled from each model's approximate posterior distribution now simulate data that are significantly visually different to the observed data.

\begin{figure}[!ht]
    \centering
\includegraphics[width=.315\textwidth]{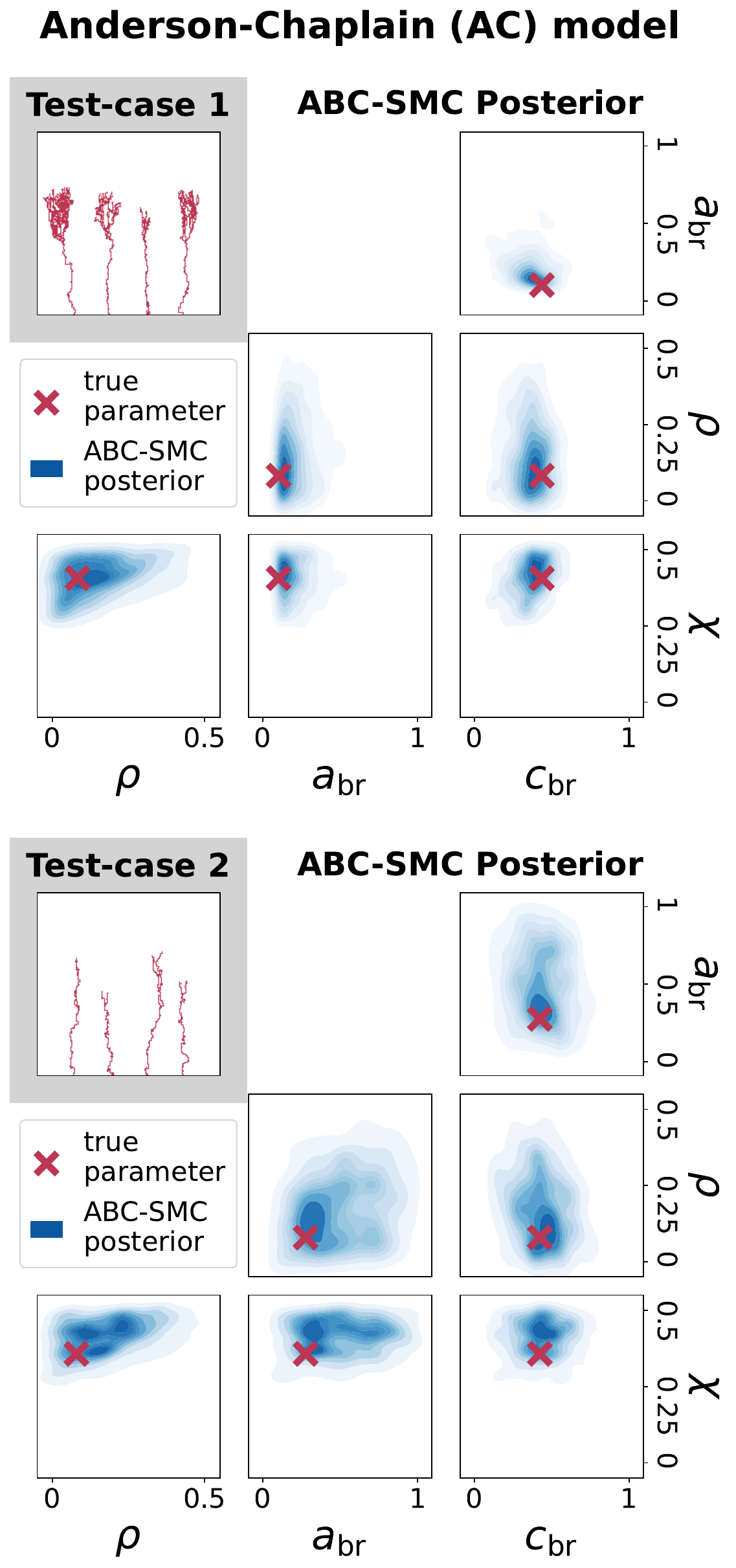}\hfill
\includegraphics[width=.315\textwidth]{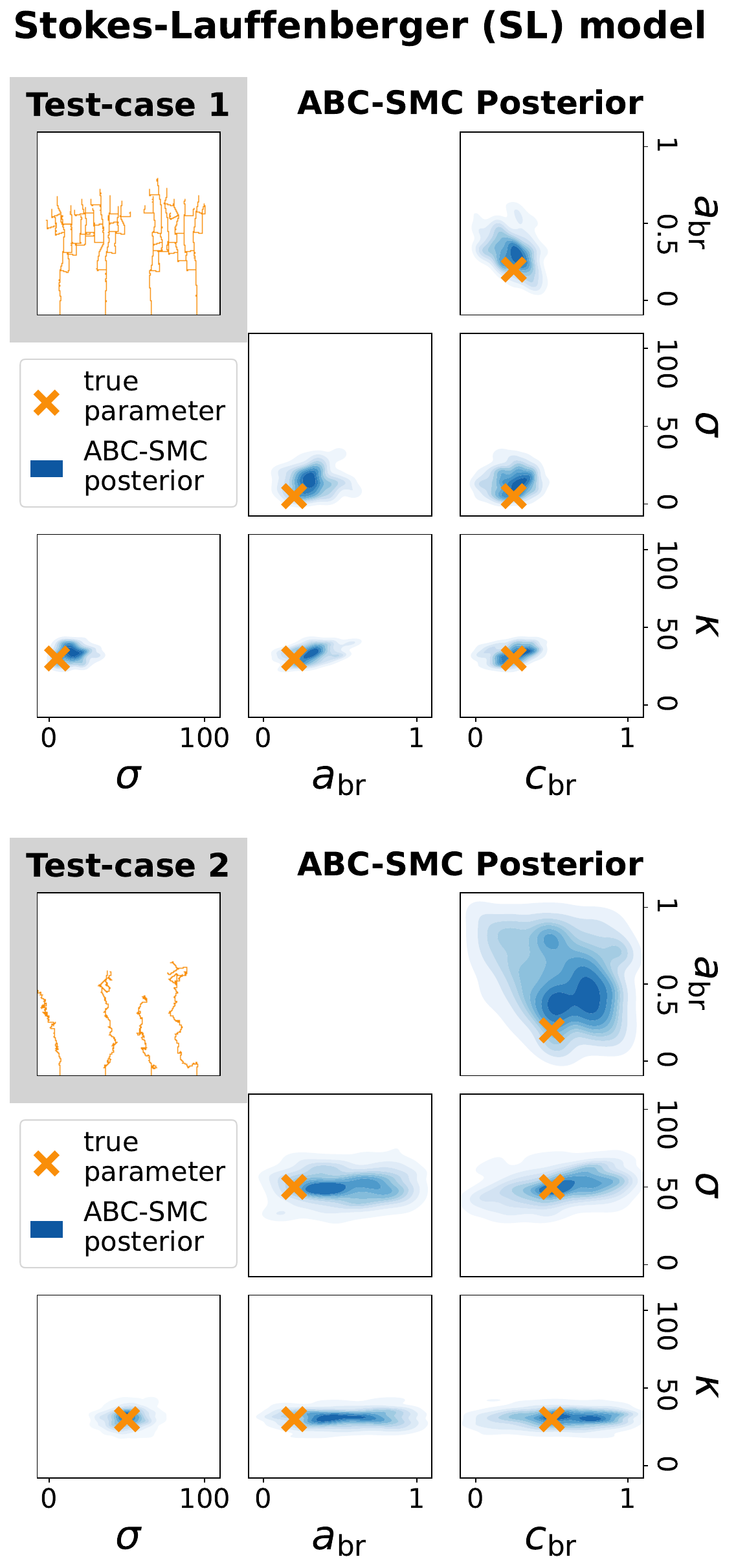}\hfill
\includegraphics[width=.315\textwidth]{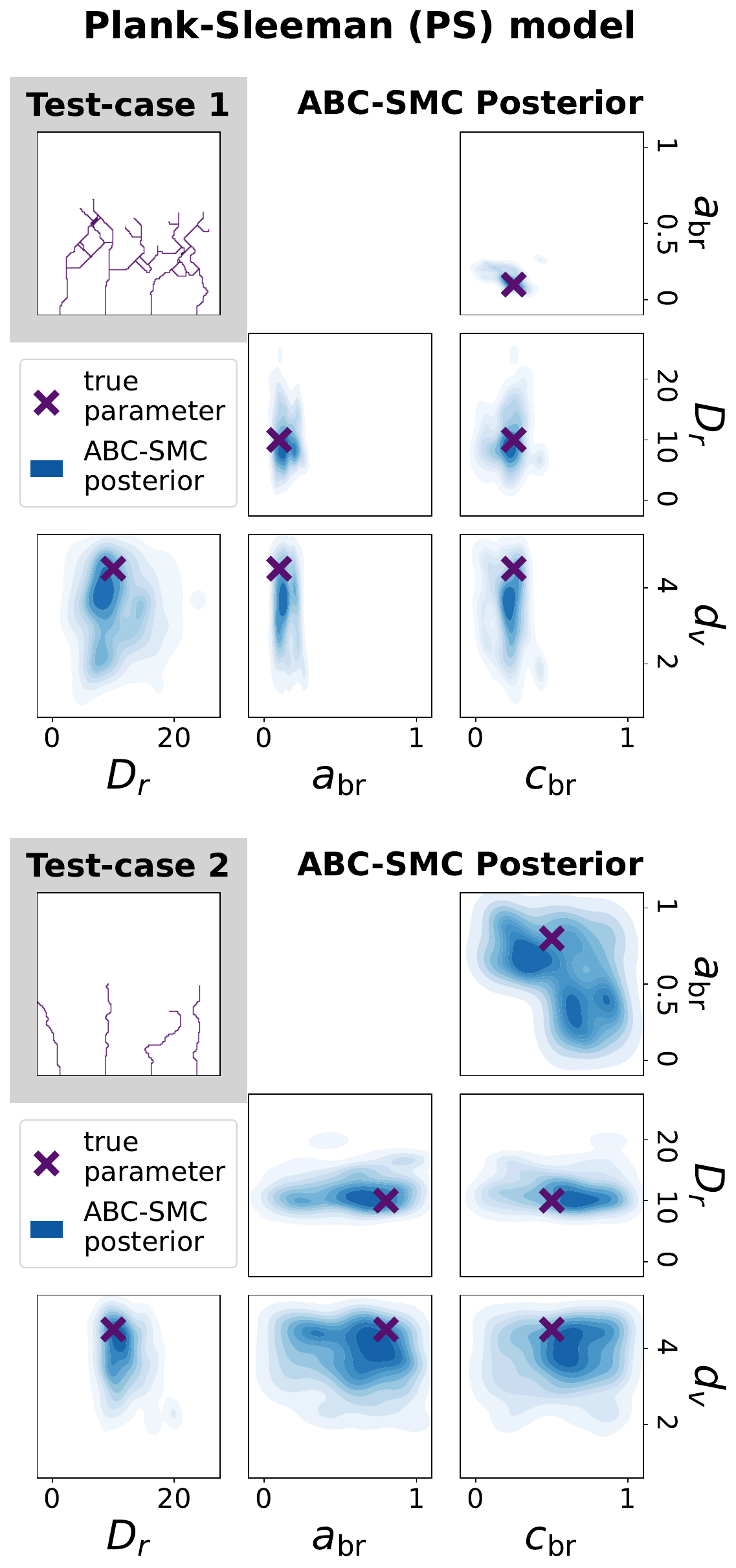}\\

    \caption{As in Figure 3 of the main text, we infer each parameter in the AC, SL, and PS models, this time modifying each model (and each test-case) to simulate EC movement for a shorter time period.
    We again simulate each model $10$ times at known parameter values to generate two synthetic test-cases for each model, and show the final time-step of one such simulation.
    We project the approximate ABC-SMC posterior to each parameter pair and plot the resulting distributions (fitting a Gaussian kernel to the parameter values accepted in the final population of the ABC-SMC algorithm).
    We also plot the true parameter which generated the test-case.
}
    \label{SI:fig:suppparaminf}
\end{figure}
\newpage

\begin{figure}[!ht]
    \centering
\includegraphics[width=.5\textwidth]{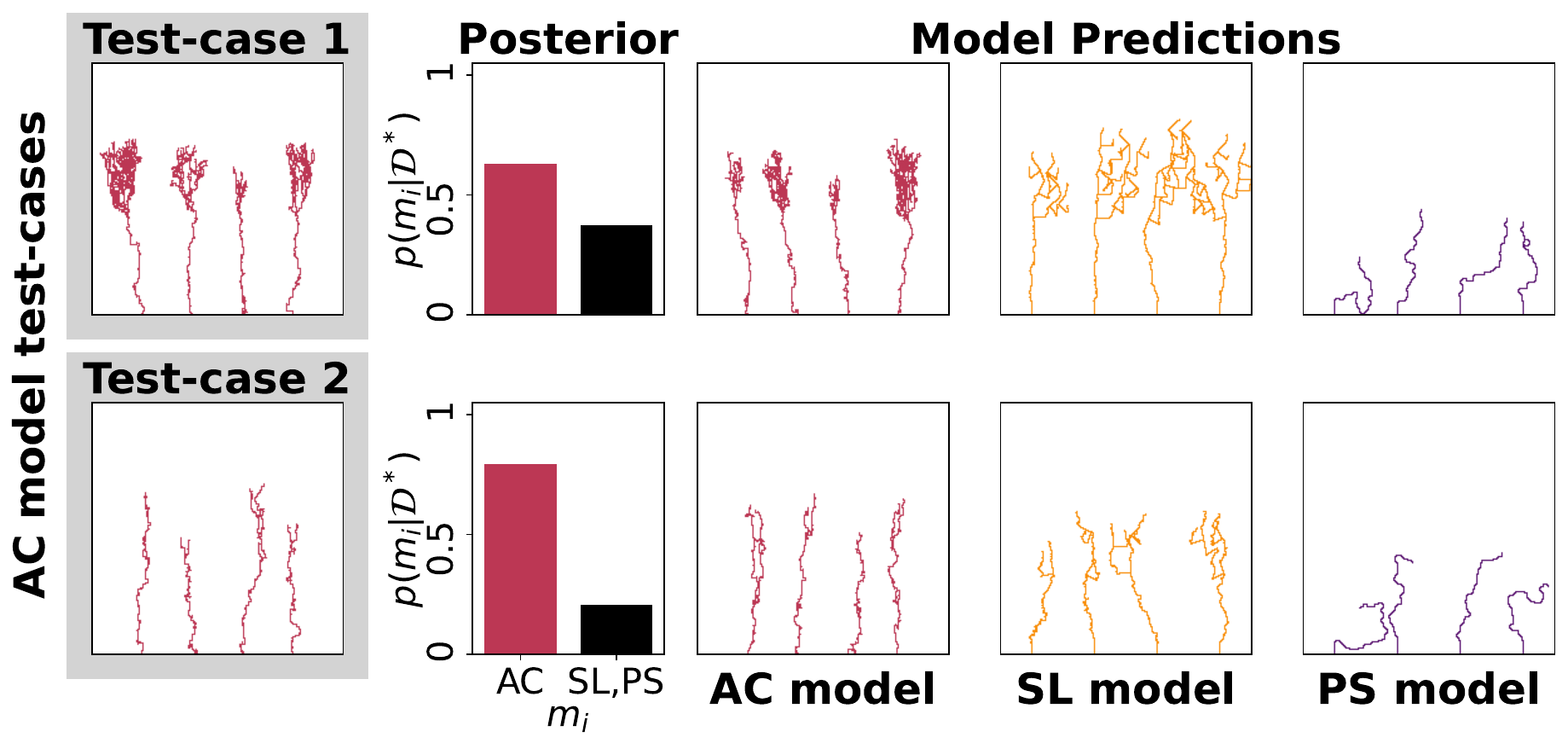}\\
\includegraphics[width=.5\textwidth]{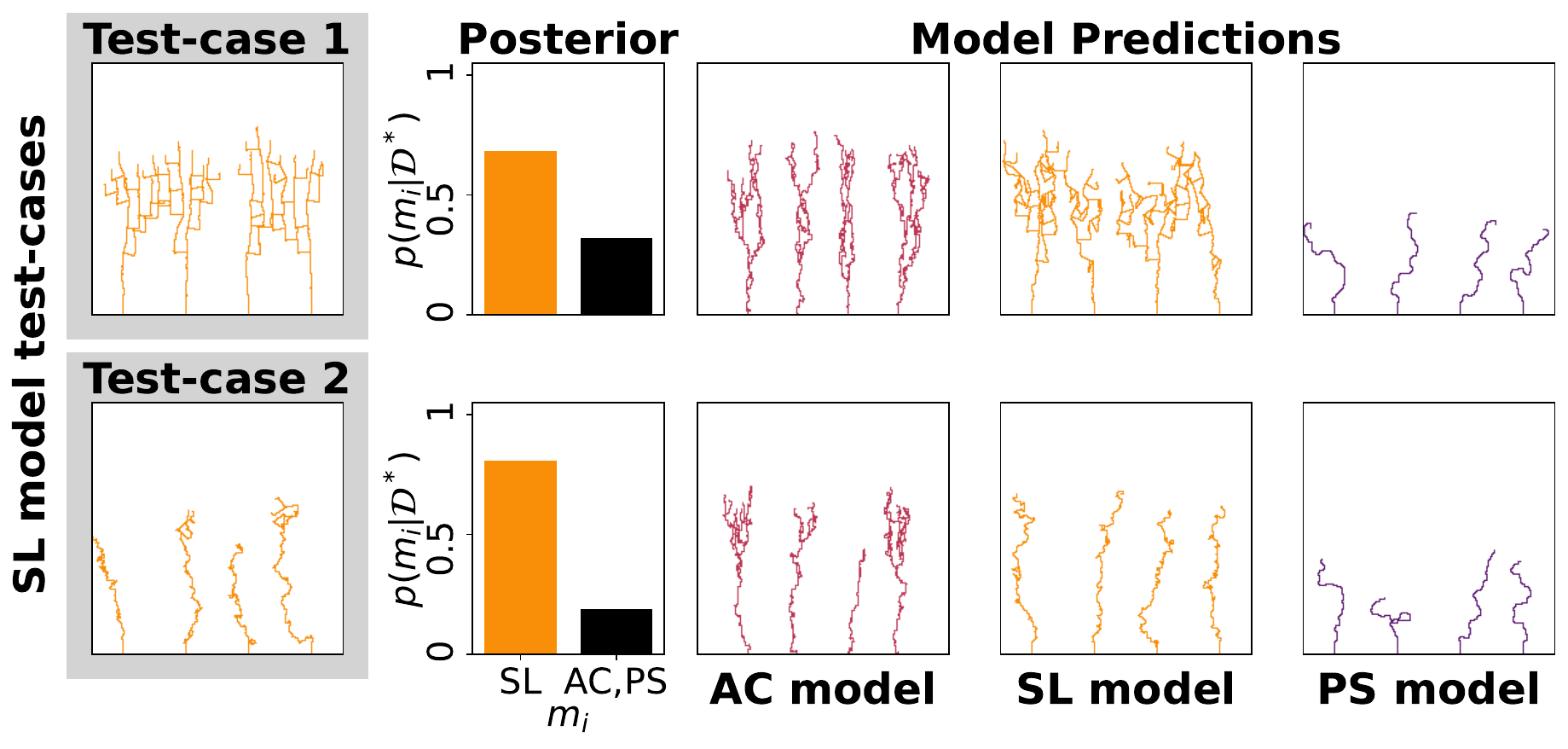}\\
\includegraphics[width=.5\textwidth]{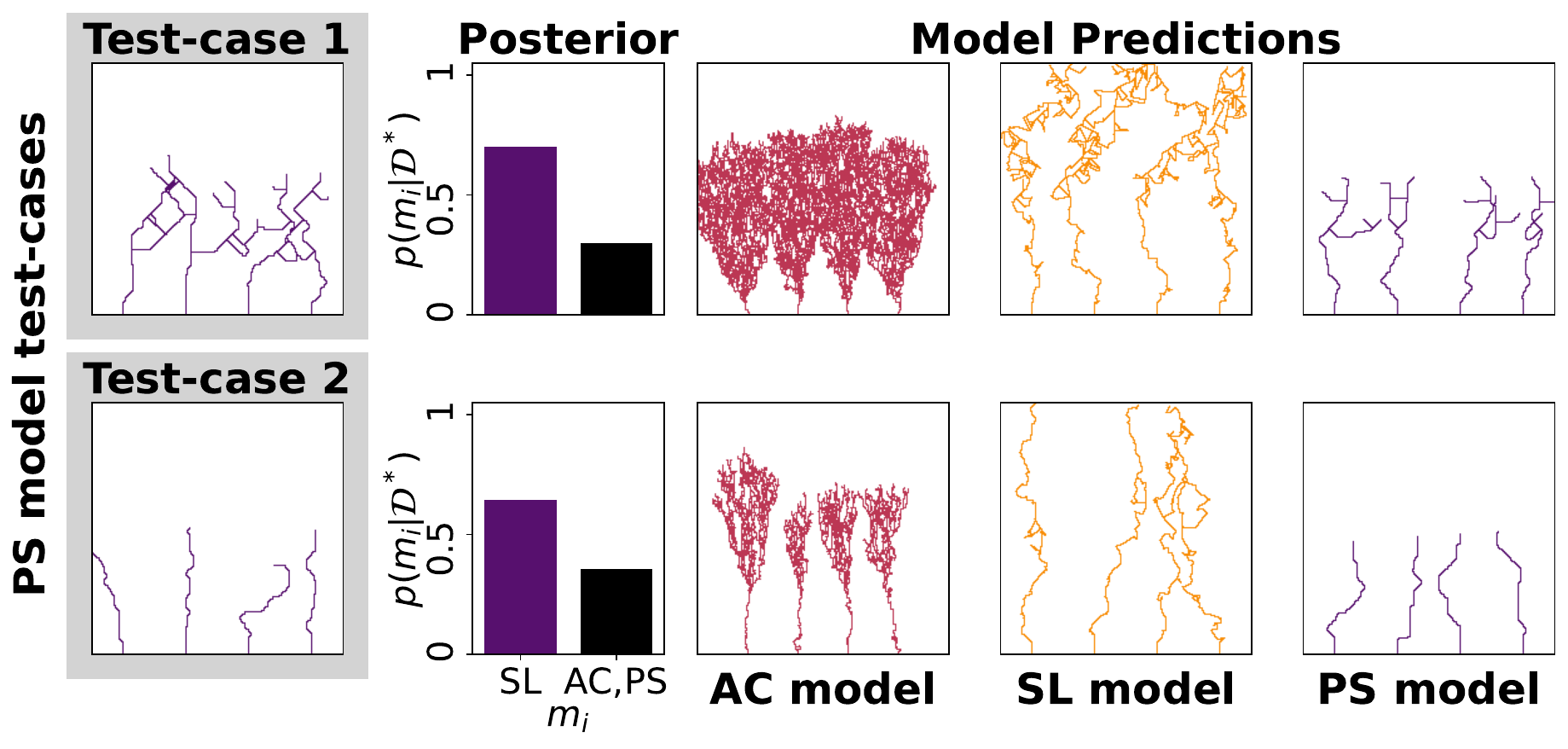}\\

    \caption{As in Figure 4 of the main text, we approximate the model posterior, again modifying each model (and each test-case) to simulate EC movement for a shorter time period.
    We again show one `prediction'-- an example of data simulated using an inferred parameter from each model's approximate parameter posterior.
    Now that some of the approximate parameter posteriors are so uncertain, some model predictions are very different from the observed data.
}
    \label{SI:fig:suppmodelselection}
\end{figure}
\newpage
    \end{appendices}
}{
}

\section*{Funding Acknowledgements}
RAM thanks the EPSRC.
HAH and HMB are grateful for the support provided by the UK Centre for Topological Data Analysis EPSRC grant EP/R018472/1. 
HAH gratefully acknowledges funding from the Royal Society RGF$\backslash$EA$\backslash$201074, UF150238 and EPSRC EP/Y028872/1 and EP/Z531224/1.
\section*{Data Availability Statement}
All code is available at \url{https://github.com/rmcdomaths/tms/} and archived at \url{https://doi.org/10.5281/zenodo.17392787}.
Parts of figures \ref{fig:importance} and \ref{fig:ph_eph} were created in Created in BioRender \url{https://BioRender.com/}.

\bibliographystyle{natbib} 
\bibliography{bib}

\end{document}